\documentclass[fleqn,usenatbib]{mnras}

\usepackage{newtxtext,newtxmath}
\usepackage[T1]{fontenc}

\DeclareRobustCommand{\VAN}[3]{#2}
\let\VANthebibliography\thebibliography
\def\thebibliography{\DeclareRobustCommand{\VAN}[3]{##3}\VANthebibliography}

\usepackage{graphicx}	% Including figure files
\usepackage{amsmath}	% Advanced maths commands

\usepackage{amssymb}	% Extra maths symbols
\usepackage{pdflscape}	% Landscape pages
\usepackage{multicol, multirow}        % Multi-column entries in tables
\usepackage{footnote}
\usepackage{ulem}

\title[Exploring the GRB population]{Exploring the GRB population: Robust afterglow modelling}

\author[M. D. Aksulu et al.]{
M. D. Aksulu,$^{1}$\thanks{E-mail: m.d.aksulu@uva.nl}
R. A. M. J. Wijers,$^{1}$
H. J. van Eerten,$^{2}$
A. J. van der Horst$^{3,4}$
\\
$^{1}$  Anton Pannekoek Institute for Astronomy, University of Amsterdam, Science Park 904, NL-1098 XH Amsterdam, the Netherlands\\
$^{2}$ Department of Physics, University of Bath, Claverton Down, Bath BA2 7AY, UK\\
$^{3}$ Department of Physics, The George Washington University, 725 21st Street NW, Washington, DC 20052, USA\\
$^{4}$ Astronomy, Physics, and Statistics Institute of Sciences (APSIS), 725 21st Street NW, Washington, DC 20052, USA\\
}

\date{Accepted XXX. Received YYY; in original form ZZZ}

\pubyear{2015}

\begin{document}
\label{firstpage}
\pagerange{\pageref{firstpage}--\pageref{lastpage}}
\maketitle

% Abstract of the paper
%@arxiver{plt-corner-t_90-epsilon_gamma-nohist.png}
\begin{abstract}
Gamma-ray bursts (GRBs) are ultra-relativistic collimated outflows, which emit synchrotron radiation throughout the entire electromagnetic spectrum when they interact with their environment. This afterglow emission enables us to probe the dynamics of relativistic blast waves, the microphysics of shock acceleration, and environments of GRBs. We perform Bayesian inference on a sample of GRB afterglow data sets consisting of 22 long GRBs and 4 short GRBs, using the afterglow model \texttt{scalefit}, which is based on 2D relativistic hydrodynamic simulations. We make use of Gaussian processes to account for systematic deviations in the data sets, which allows us to obtain robust estimates for the model parameters. We present the inferred parameters for the sample of GRBs, and make comparisons between short GRBs and long GRBs in constant-density and stellar-wind-like environments. We find that in
almost all respects such as energy and opening angle, short and long GRBs
are statistically the same. Short GRBs however have a markedly lower 
prompt $\gamma$-ray emission efficiency than long GRBs. We also find that
for long GRBs in ISM-like ambient media there is a significant anti-correlation between the fraction of thermal energy in the magnetic fields, $\epsilon_B$, and the beaming corrected kinetic energy. 
Furthermore, we find no evidence that the mass-loss rates of the progenitor
stars are lower than those of typical Wolf-Rayet stars.

\end{abstract}

% Select between one and six entries from the list of approved keywords.
% Don't make up new ones.

%%%%%%%%%%%%%%%%%%%%%%%%%%%%%%%%%%%%%%%%%%%%%%%%%%

%%%%%%%%%%%%%%%%% BODY OF PAPER %%%%%%%%%%%%%%%%%%

\section{Introduction}
Gamma-ray bursts (GRBs) are the most powerful explosions in the Universe. They are ultra-relativistic collimated outflows, which are powered by a compact central object. GRBs are initially observed as brief flashes of $\gamma$ rays lasting about 0.1--1000\,s. These initial brief flashes of high-energy radiation are called the prompt emission of the GRB. The exact emission mechanism of the prompt emission remains elusive, despite decades of dedicated research. GRBs are phenomenologically categorized as short and long GRBs depending on the observed duration of the prompt emission phase \citep{1993ApJ...413L.101K}. Short GRBs have been associated with compact object mergers where at least one of the objects is a neutron star \citep{LattimerSchramm1976,1989Natur.340..126E}, whereas long GRBs are thought to be results of core-collapse supernovae of massive stars \citep{1993ApJ...405..273W}.

As the ejected ultra-relativistic outflow from a GRB starts to interact with the circumburst medium (CBM), a pair of shocks are generated, one of which propagates into the ejecta (reverse shock) and the other propagates into the CBM (forward shock). In these shocks, tangled magnetic fields are amplified and charged particles are accelerated, which results in long-lasting synchrotron emission spanning the whole electromagnetic spectrum \citep{1992MNRAS.258P..41R}. This broadband synchrotron emission is observable for several months, even years in some cases, and is called the afterglow emission of the GRB. The afterglow emission provides crucial insights on the energetics and environments of GRBs, the dynamics of relativistic blast waves, and the microphysics of particle acceleration in shocks \citep{1997MNRAS.288L..51W, 1998ApJ...497L..17S, 1999ApJ...523..177W, 2002ApJ...571..779P, 2003ApJ...597..459Y}.

Thanks to missions like the \textit{Neil Gehrels Swift Observatory} \citep{2004ApJ...611.1005G} and \textit{Fermi Gamma-ray Space Telescope} \citep{2009ApJ...697.1071A}, the number of detected/localized GRBs has increased and allowed for rapid, ground/space based, broadband follow-up observations of the afterglow emission. Moreover, the start of the multi-messenger era has supplemented our understanding of the physics of GRBs \citep{2017PhRvL.119p1101A, 2019Natur.575..455M}. Besides advances in observational instruments, developments in numerical hydrodynamics and radiative transfer have enabled us to build models with increasing complexity and accuracy (e.g., \citealt{2012ApJ...749...44V, 2012ApJ...751...57D, 2013ApJ...767..141V, 2015ApJ...799....3R, 2018ApJ...865...94D, 2018ApJ...869...55W, 2021MNRAS.tmp..938J}). Moreover, advances in statistical methods allow us to perform robust Bayesian inference and obtain reliable parameter estimates \citep{2020MNRAS.497.4672A}. Due to all these developments, we can now  model a sample of GRB afterglow data sets, consistently, and investigate the distribution of physical parameters in the GRB population.

Previously, \cite{2002ApJ...571..779P, 2003ApJ...597..459Y} have performed broadband afterglow modelling, and inferred burst parameters for a sample of long GRBs. Furthermore, \cite{2015ApJ...815..102F} have gathered data for a large number of short GRBs, and inferred their burst parameters based on their afterglow emission. These studies utilized semi-analytic models to reproduce the observed broadband emission; in this study we make use of a model based on 2D relativistic hydrodynamic simulations. This allows us to capture the dynamics of these energetic events in a more realistic fashion.

In \cite{2020MNRAS.497.4672A} (A20, from now on), we introduced a new method for Bayesian parameter estimation, where we make use of Gaussian processes (GPs) in order to take into account some systematic effects in the data set and physics not included in the model. We showed in A20 that this approach allows us to obtain more robust parameter estimates, whereas the more conventional method of sampling the $\chi^2$ likelihood leads to underestimated uncertainties on the parameters, especially in the presence of systematics. We make use of a modified version of the GP model described in A20, in order to model a sample of 26 GRB afterglow data sets. In Section \ref{sec:method} we describe the GRB sample, model, inference approach and details of the regression process. In Section \ref{sec:results} we present our results and the inferred physical parameters of the sample. Finally we discuss our findings in Section \ref{sec:discussion} and conclude in Section \ref{sec:conclusion}. Throughout this work we assume the cosmology as described in \cite{2016A&A...594A..13P}.

%----------------------
\section{Method}
\label{sec:method}
\subsection{Sample}
\label{sec:sample}

\begin{table}
    \caption{The GRB sample for this study. The measured redshift ($z$) and isotropic equivalent prompt energetics ($E{\gamma, \mathrm{iso}}$) are presented.}
    \label{tab:grb_sample}
    \begin{tabular} {|c|l|r@{.}l|r@{.}l@{$\ \pm\ $}r@{.}l|}
        \hline
        \multicolumn{2}{c}{Burst name} &  \multicolumn{2}{c}{$z$} & \multicolumn{4}{c}{$E{\gamma, \mathrm{iso}} / 10^{52}$ (erg)} \\ 
        \hline
        \hline
        \parbox[t]{2mm}{\multirow{4}{*}{\rotatebox[origin=c]{90}{short GRBs}}} & 051221A & 0&5465 & $0$&\multicolumn{3}{@{}l}{$15$} \\
        & 130603B & 0&3564 & $ 0$&\multicolumn{3}{@{}l}{$21$} \\
        & 140903A & 0&351 & $ 0$&$006$&$0$&$0003$   \\
        & 200522A & 0&5536 & $ 0$&$0084$&$0$&$0011$   \\
        \hline
        \hline
        \parbox[t]{2mm}{\multirow{22}{*}{\rotatebox[origin=c]{90}{long GRBs}}} & 970508 & 0&835 & $0$&$61$&$0$&$13$ \\
        & 980703 & 0&966 & $6$&$9$&$0$&$8$  \\
        & 990510 & 1&619 & $17$&$8$&$2$&$6$ \\
        & 991208 & 0&706 & $22$&$3$&$0$&$8$ \\
        & 991216 & 1&02 & $ 67$&$5$&$8$&$1$ \\
        & 000301C & 2&04 & $4$&\multicolumn{3}{@{}l}{$6$}\\
        & 000418 & 1&118 & $9$&$1$&$1$&$7$  \\
        & 000926 & 2&066 & $27$&&$5$&$8$  \\
        & 010222 & 1&477 & $81$&&$1$   \\
        & 030329 & 0&1685 & $1$&$66$&$0$&$2$  \\
        & 050820A & 2&615 & $97$&$5$&$7$&$7$   \\
        & 050904 & 6&29 & $124$&&$7$&$7$   \\
        & 060418 & 1&49 & $12$&$8$&$1$&   \\
        & 090328 & 0&7357 & $13$&&$3$&   \\
        & 090423 & 8&26 & $9$&$5$&$2$&   \\
        & 090902B & 1&8229 & $440$&&$30$&   \\
        & 090926A & 2&1062 & $200$&&$5$&   \\
        & 120521C & 6&0 & $8$&$25$&$2$&  \\
        & 130427A & 0&3399 & $81$&\multicolumn{3}{@{}l}{}\\
        & 130702A & 0&145 & $0$&$064$&$0$&$01$   \\
        & 130907A & 1&238 & $330$&&$10$&   \\
        & 140304A & 5&283 & $12$&$24$&$1$&$4$   \\
        \hline
        \hline
    \end{tabular}
\end{table}

Our GRB afterglow sample consists of 26 GRBs with well-sampled, broadband data sets. We relied only on peer-reviewed, published data sets, and converted the reported measurements to mJy units. The main selection criterion for the sample of GRBs has been the availability of broadband afterglow data. 22 out of the 26 GRBs are long GRBs detected between 1997--2014, with published broadband data sets; the time period is set to get a large enough sample. For short GRBs, we found only four with detections in radio, optical and X-ray bands up to the present. We omitted GRBs with non-canonical features in their light curves and include the five GRBs modelled in A20. When possible, we neglect epochs and/or bands for which there is evidence that the emission is dominated by processes which are not included in our model (e.g., early time optical and radio emission from GRB 130427A, which is dominated by reverse shock emission). This does not, of course, in  any way represent a well-defined complete 
sample. We drew the boundary for having enough data somewhat subjectively, and similarly selected data sections
in the early light curves suspected of unmodeled physics by eye.

We corrected the observed flux values for Galactic dust extinction using the extinction curve given by \cite{1992ApJ...395..130P}. We subtract any persistent emission originating from the host galaxy when possible. We do not correct the data for the dust extinction due to the host galaxy; instead we leave the rest-frame $A_V$ value for the host galaxy as a free parameter (see Section \ref{sec:model}).  
We present the GRB sample in Table \ref{tab:grb_sample}.

\subsection{Gaussian process framework}

GPs are stochastic processes which can be used for regression and classification problems for which the underlying physical model is unknown (e.g., \citealt{2006gpml.book.....R}). Following A20 (also see \citealt{2012MNRAS.419.2683G}), we make use of GPs to take into account any systematic deviations from the afterglow model. In this section, we highlight some improvements on the GP model introduced in A20. For clarity we use the same notation as in A20. Vectors and matrices are represented by bold symbols.

The systematics are described by the GP model as,
\begin{equation}
    f\left(t, \nu\right) \sim \mathcal{GP}\left(\mu(t, \nu, \boldsymbol{\phi}), \boldsymbol{\Sigma}(t, \nu, \boldsymbol{\theta})\right),
\end{equation}
where $t$ and $\nu$ are the time and frequency coordinates in the observer frame, $\boldsymbol{\phi}$ represents the afterglow model parameters, and $\boldsymbol{\theta}$ represents the hyperparameters of the GP. Since the observer time and frequency change over many orders of magnitude, we work with the logarithm of these coordinates when performing GP regression. The mean function of the GP, $\mu$, is the afterglow model, and $\boldsymbol{\Sigma}$ is the covariance matrix which describes how the systematics are correlated over $t$ and $\nu$. We adopt a 2D heterogeneous squared-exponential kernel function (e.g., see \citealt{2006gpml.book.....R}) to calculate the covariance matrix,
\begin{equation}
    \boldsymbol{\Sigma}_{ij} = k(\boldsymbol{\mathrm{X}}_i, \boldsymbol{\mathrm{X}}_j) = A \exp\left[-\frac{1}{2}\sum^2_{k = 1}\frac{(\boldsymbol{\mathrm{X}}_{ik}- \boldsymbol{\mathrm{X}}_{jk})^2}{l_k^2}\right] + \delta_{ij}\sigma_h^2,
\end{equation}
where $\boldsymbol{\mathrm{X}}$ represents the 2D feature set (i.e., observer time and band). The hyperparameters of the GP are defined as,
\begin{equation}
    \boldsymbol{\theta} = (A, l_1, l_2, \sigma_h)^T
    \label{eq:hyperparams}
\end{equation}
\noindent where $A$ represents the amplitude of the correlations,  $l_1$ and $l_2$ determine the length scales of the correlations over time and frequency, respectively, and $\sigma_h$ represents the amount of white noise in the data set. In A20, the systematics in the data set were assumed to be uncorrelated across different observational bands. Therefore, the hyperparameter $l_2$ was fixed to be a small number. In this work, we make $l_2$ a free parameter, thereby allowing the GP model to capture systematics correlated over different observational bands. We refer the reader to A20 for a more detailed explanation of the GP model.

\subsection{Model}
\label{sec:model}
We assume a collimated, ultra-relativistic blast wave moving into a circumburst medium, which need not be uniform; we assume a  density profile of the form
\begin{equation}
    n=n_{\mathrm{ref}} \left(\frac{r}{10^{17}\ \mathrm{cm}}\right) ^{-k}.
    \label{eq:densityprofile}
\end{equation} 
The normalisation is chosen at a radius that often falls within the range sampled by real afterglows. We will not treat
$k$ as a fully free parameter, but only allow the `classic' values of 2 and 0. For $k=2$, the interpretation of the
environment is pretty unambiguous: the blast wave is in the unshocked, freely expanding part of a massive stellar wind.
In that case a more common nomenclature of the parameters is:
\begin{equation}
    \rho(r) = \frac{A}{r^2},
\end{equation}
where $r$ is the distance from the star and $A$ can be expressed as,
\begin{equation}
    A = \frac{\dot{M}}{4\pi v_{\mathrm{wind}}}.
\end{equation}
Here, $\dot{M}$ is the mass loss rate of the progenitor star and $v_{\mathrm{wind}}$ is the wind velocity. For a canonical Wolf-Rayet star with a mass loss rate of $10^{-5}\,M_\odot / \mathrm{yr}$ and wind velocity of $1000\,\mathrm{km / s}$, $A\sim5\times 10^{11}\,\mathrm{g / cm}$, which value is denoted by $A_*$.  We can simply scale the inferred $n_{\mathrm{ref}}$ values to units of $A_*$ using
\begin{equation}
   \frac{A}{A_*} = \frac{n_{\mathrm{ref}}}{30\,\mathrm{cm^{-3}}}.
\end{equation}
For $k=0$ the interpretation of the environment is much more ambiguous. In this case the actually observed 
afterglow typically covers well under a factor 10 in radius travelled by the blast wave, so any environment in which
the density does not change much over a factor few in distance (and within the solid angle hit by the outflow) will do.
This could definitely be canonical ISM, but for $A_*$ not too different from 1, a wind bubble around a massive star 
will contain many solar masses of material and thus the GRB jet will never emerge from it during the normal afterglow phase (the blast wave typically needs to sweep up less than (beamed equivalent of a spherical amount of)  0.1\,M$_\odot$ of ambient matter to become non-relativistic). However, the bulk of the wind bubble will contain wind that has been shocked against the ISM, and that is
uniform enough to fit the $k=0$ case. Another possibility might be that the star has a significant proper motion through
a somewhat dense ISM. In that case most of the wind bubble is swept back, and in the forward hemisphere the blast wave may emerge from the wind into the ISM in time.

The initial Lorentz factor of the blast wave is assumed to be uniform within the opening angle of the jet, i.e. a top hat jet model. We assume that charged particles are accelerated in the forward shock and emit synchrotron emission \citep{1998ApJ...497L..17S, 1999ApJ...523..177W, 2002ApJ...568..820G}. In this work, we do not take into account emission originating from the reverse shock and thus confine ourselves to fitting the later parts of the afterglow when the reverse shock has passed through the ejecta and the deceleration phase is over; in this limit the value of the initial Lorentz factor of the jet is no longer important and need not (indeed, cannot) be fit. For the `microphysical' parameters, which describe the spectrum and energy content of the electrons behind the blast wave and the magnetic field in which they move, we use the customary notation: $p$ is the power-law index of the energy distribution
of the relativistic electrons, and $\epsilon_e$ and $\epsilon_B$ are the fractions of post-shock
energy density in relativistic electrons and magnetic field, respectively. Only a fraction of all electrons, $\xi_N$, may
be accelerated. When $p\simeq2$, the total energy in electrons and the value
of $p$ become very correlated in the fit, because the blast-wave emission depends on the combination $\Bar{\epsilon}_e \equiv \frac{p - 2}{p - 1}\epsilon_e$. Therefore we fit for that quantity, and disentangle $p$ and
$\epsilon_e$ later where possible. Similarly, the fraction of accelerated electrons, $\xi_N$, is degenerate with respect to  $(E_{K,\mathrm{iso}}, n_\mathrm{ref}, \epsilon_B, \Bar{\epsilon}_e)$, where $(E_{K,\mathrm{iso}}, n_\mathrm{ref})$ are proportional to $1 / \xi_N$, and $(\epsilon_B,  \bar{\epsilon}_e)$ are proportional to $\xi_N$ \citep{2005ApJ...627..861E}. Because of this degeneracy, we cannot determine $\xi_N$ independently from afterglow light curves and fix it to the canonical value of 1 (for ease of comparison with previous studies).

We make use of the numerical model \texttt{scalefit} \citep[Ryan et al. in preparation;][]{2020MNRAS.497.4672A, 2015ApJ...799....3R}. \texttt{scalefit} uses pre-calculated tables of spectral features (spectral breaks, peak spectral flux) for a range of different time epochs, opening angles, and observing angles. These tables are generated separately for ISM and wind-like circumburst density profiles using \texttt{boxfit} \citep{2012ApJ...749...44V}. \texttt{boxfit} is a numerical code which is able to output flux values for given observer time, frequency, and GRB parameters. The main advantage of \texttt{boxfit} is that the dynamics rely on pre-calculated relativistic hydrodynamic (RHD) simulations. However, since \texttt{boxfit} solves the radiative transfer equations during runtime, it is computationally expensive. Therefore, it is not practical to use \texttt{boxfit} when performing Bayesian inference. Moreover, \texttt{boxfit} does not take into account the effects of synchrotron cooling on the self-absorption break. This may lead to incorrect spectra in certain regimes. \texttt{scalefit}, on the other hand, makes use of pre-calculated spectral features, obtained from \texttt{boxfit} in a valid regime, and utilizes scaling rules \citep{2012ApJ...747L..30V} to calculate the spectra for various regimes (i.e. different orderings of the break frequencies). \texttt{scalefit} is valid for all spectral regimes, unlike \texttt{boxfit}, and is computationally inexpensive in comparison \citep[Ryan et al. in preparation][]{}. However, \texttt{scalefit} makes assumptions about the sharpness of the spectra around break frequencies, whereas \texttt{boxfit} generates smooth spectra in a self-consistent way.

Additionally, we account for dust extinction due to the host galaxy when calculating the observed flux. For the majority of GRBs in our sample we adopt the Small Magellanic Cloud extinction curve given by \cite{1992ApJ...395..130P}. However, for GRBs 000418 \citep{2003A&A...409..123G}, 010222 \citep{2002ApJ...565..829F} and 090328 \citep{2010A&A...516A..71M} we assume a Starburst type extinction curve \citep{2000ApJ...533..682C}. We include the dust extinction due to the host galaxy as a free parameter. 

Summarizing, our model parameters are defined as
\begin{equation}
    \boldsymbol{\phi} = (\theta_0, E_{K, \mathrm{iso}}, n_\mathrm{ref}, \theta_{\mathrm{obs}}, p, \epsilon_B, \Bar{\epsilon}_e, \xi_N, A_V)^T,
    \label{eq:model_params}
\end{equation}
\noindent where $\theta_0$ is the opening angle of the jet, $E_{K, \mathrm{iso}}$ is the isotropic-equivalent kinetic energy of the explosion, $n_\mathrm{ref}$ is the normalization factor for the circumburst density profile (see Equation \ref{eq:densityprofile}), $\theta_{\mathrm{obs}}$ is the observing angle, $p$ is the power-law index of the accelerated electron population, $\epsilon_B$ is the fraction of post-shock energy in the magnetic fields, $\Bar{\epsilon}_e \equiv \frac{p - 2}{p - 1}\epsilon_e$ where $\epsilon_e$ is the fraction of post-shock energy in the accelerated electrons, $\xi_N$ is the fraction of electrons being accelerated, and $A_V$ is the amount of dust extinction in the rest-frame due to the host galaxy. 

\subsection{Regression}
In order to obtain posterior distributions for the hyperparameters and model parameters, we make use of nested sampling \citep{2004AIPC..735..395S}. Incorporating nested sampling allows us to calculate the evidence with an associated numerical uncertainty, while producing posterior samples as a byproduct. Inferring the Bayesian evidence is instrumental in this study, because it gives us a measure to determine which model explains the data best: a blast wave moving into a homogeneous ($k=0$) or wind-like ($k=2$) circumburst medium (see Section~\ref{sec:model}).

Following A20, we utilize \texttt{pymultinest} \citep{2014A&A...564A.125B}, which is a \texttt{PYTHON} package based on the MultiNest nested sampling algorithm \citep{2009MNRAS.398.1601F}. For all the presented results, \texttt{pymultinest} is used in the importance sampling mode \citep{2019OJAp....2E..10F} with mode separation disabled. We use 400 initial live points and use an evidence tolerance of 0.5 as our convergence criterion. 

We assume wide priors for $a$ and $\sigma_h$, however, the length scale hyperparameters (i.e. $l_1$ and $l_2$) are capped at 1 (see Table~\ref{tab:prior_hp}), since we do not expect any systematics to be correlated over orders of magnitude (the GP model operates in the log-space). This is important, we found, because if one allows long correlation length scales, the GP can take up features like constant offsets between model and data, or slope differences, which the model should really be capable of fitting. We intend the Gaussian process mostly to take up issues like calibration differences between instruments leading to extra 'noise' within a band, and physical effects that are shorter in time and frequency scale than is included in the model,
such as radio scintillation, minor flares, etc.

For all the model parameters we assume uninformative prior distributions, which can be seen in Table~\ref{tab:prior_model}.

Note that we do not take into account any reported upper limits on the afterglow flux when inferring parameters, since upper limit reports typically do not contain enough information to include them in the fitting in a statistically sound way.

\begin{table}
    \setlength{\tabcolsep}{5pt}
    \def\arraystretch{1.25}
    \caption{Assumed priors for the GP hyperparameters.}
    \begin{tabular}{l|r}
    \hline
    \multicolumn{1}{l}{Parameter range} & \multicolumn{1}{r}{Prior distribution} \\
    \hline
    $10^{-10}<a<10^{10}$ & log-uniform\\
    
    $10^{-6}<l_1<1$ & log-uniform\\
    
    $10^{-6}<l_2<1$ &log-uniform\\
    
    $10^{-3}<\sigma_h<10^{3}$ & log-uniform\\
    \hline
    \end{tabular}
    \label{tab:prior_hp}
\end{table}
\begin{table}
    \setlength{\tabcolsep}{5pt}
    \def\arraystretch{1.25}
    \caption{Assumed priors for the physical parameters.}
    \begin{tabular}{l|r}
    \hline
    \multicolumn{1}{l}{Parameter range} & \multicolumn{1}{r}{Prior distribution} \\
    \hline
    $0.01<\theta_0<1.6$ & log-uniform\\
    
    $10^{50}<E_{K, \mathrm{iso}}<10^{56}$ & log-uniform\\
    
    $10^{-3}<n_{\mathrm{ref}}<1000$ &log-uniform\\
    
    $0<\theta_{\mathrm{obs}} / \theta_0<2$ & uniform\\
    
    $1.0<p<3.0$& uniform\\
    
    $10^{-10}<\epsilon_B<1.0$& log-uniform\\
    
    $10^{-10}<\bar{\epsilon_e}<10$& log-uniform\\
    
    $0<A_V<10$& uniform\\
    \hline
    \end{tabular}
    \label{tab:prior_model}
\end{table}

\section{Results}
\label{sec:results}

\begin{table}
    \setlength{\tabcolsep}{5pt}
    \def\arraystretch{1.25}
    \caption{Model selection for the GRB sample. $\mathcal{Z}$ represents the Bayesian evidence. Reported uncertainties represent $1$--$\sigma$. We select the mode, either homogeneous (ISM) or wind-like environment, with higher inferred evidence values. The evidence values for the preferred model are written in \textbf{bold} numerals.  }
    \begin{tabular}{|c|l|r|r|r@{.}l|}
    \hline
    \multicolumn{1}{l}{} & \multicolumn{1}{l}{Burst name} & \multicolumn{1}{r}{$\ln\mathcal{Z}$ [ISM]} & \multicolumn{1}{r}{$\ln\mathcal{Z}$ [Wind]} & \multicolumn{2}{r}{Bayes factor}\\
    \hline
    \hline
    \parbox[t]{2mm}{\multirow{4}{*}{\rotatebox[origin=c]{90}{short GRBs}}} & 051221A & $\mathbf{-22.79 \pm 0.04}$ & $-24.26 \pm 0.12$& $4$&$33$ \\
    & 130603B$^{(\mathrm{a})}$ & $-21.45 \pm 0.05$ & $-21.42 \pm 0.03$ & $\sim1$&\\
    & 140903A & $\mathbf{-24.97 \pm 0.05}$ & $-25.73 \pm 0.02$  & $2$&$15$ \\
    & 200522A & $\mathbf{-18.31 \pm 0.03}$ & $-19.30 \pm 0.02$  & $2$&$68$ \\
    \hline
    \hline
    \parbox[t]{2mm}{\multirow{22}{*}{\rotatebox[origin=c]{90}{long GRBs}}} & 970508 & $-99.21 \pm 0.12$ & $\mathbf{-92.68 \pm 0.31}$ & $>150$& \\
    & 980703 & $-86.74 \pm 0.02$ & $\mathbf{-82.33 \pm 0.06}$ & $82$&$80$\\
    & 990510 & $\mathbf{279.21 \pm 0.02}$ & $278.45 \pm 0.03$ & $2$&$15$\\
    & 991208 & $\mathbf{-60.33 \pm 0.04}$ & $-67.63 \pm 0.10$ & $>150$&\\
    & 991216 & $-5.37 \pm 0.04$ & $\mathbf{-4.56 \pm 0.03}$ & $2$&$25$\\
    & 000301C & $\mathbf{37.45 \pm 0.05}$ & $25.30 \pm 0.12$ & $>150$&\\
    & 000418 & $-55.09 \pm 0.04$ & $\mathbf{-49.81 \pm 0.05}$ & $>150$&\\
    & 000926 & $28.31 \pm 0.08$ & $\mathbf{34.48 \pm 0.04}$ & $>150$&\\
    & 010222 & $\mathbf{37.34 \pm 0.04}$ & $31.01 \pm 0.02$ & $>150$&\\
    & 030329 & $\mathbf{-29.23 \pm 0.01}$ & $-59.81 \pm 0.05$ & $>150$&\\
    & 050820A & $-40.26 \pm 0.74$ & $\mathbf{-33.88 \pm 0.05}$ & $>150$&\\
    & 050904 & $\mathbf{-31.20 \pm 0.03}$ & $-33.30 \pm 0.07$ & $8$&$13$\\
    & 060418 & $\mathbf{-11.55 \pm 0.06}$ & $-19.19 \pm 0.02$ & $>150$&\\
    & 090328 & $\mathbf{-50.14 \pm 0.03}$ & $-51.69 \pm 0.30$ & $4$&$71$\\
    & 090423 & $\mathbf{-51.42 \pm 0.06}$ & $-55.97 \pm 0.10$ & $94$&$59$\\
    & 090902B & $-49.39 \pm 0.02$ & $\mathbf{-39.78 \pm 0.04}$ & $>150$&\\
    & 090926A & $\mathbf{-9.68 \pm 0.03}$ & $-12.24 \pm 0.02$ & $12$&$98$\\
    & 120521C & $\mathbf{-54.96 \pm 0.06}$ & $-55.50 \pm 0.09$ & $1$&$70$\\
    & 130427A & $324.52 \pm 0.08$ & $\mathbf{336.86 \pm 0.03}$ & $>150$&\\
    & 130702A & $\mathbf{19.45 \pm 0.18}$ & $8.55 \pm 0.68$ & $>150$&\\
    & 130907A & $\mathbf{-135.85 \pm 0.01}$ & $-141.59 \pm 0.02$ & $>150$&\\
    & 140304A & $-60.46 \pm 0.04$ & $\mathbf{-57.90 \pm 0.04}$ & $13$&$00$\\
    \hline
    \hline
    \end{tabular}
    \begin{flushleft}
    \begin{small}

     (a) For this data set both homogeneous and wind-like models result in similar evidence values. An ISM-type environment is preferred since this is a short GRB and no strong winds are expected due their progenitors.
    \end{small}
    \end{flushleft}
    \label{tab:evidence}
\end{table}

In this section, we present the modelling results for our sample of 26 GRB afterglow data sets (see Table \ref{tab:grb_sample}). We will not discuss individual GRBs in detail, since the objective of our work is to 
examine the properties of a population of GRB afterglow sources and systematics of how the properties are
distributed and may differ between subclasses. In so doing, we will examine correlations between each pair of fit parameters and distributions of fit parameters between each of a few subclasses. All in all, we make about 50
such comparisons, and therefore we have a fair chance of finding differences or correlations at the few percent
probability level by statistical coincidence. To account for this, we will only regard correlations or differences
in distributions as firmly significant when the null hypothesis of no correlation or no difference can be excluded
at the single-trial $p$ value of $3\times10^{-4}$ or better, and tentative below $p=1\times10^{-3}$. Of course, since
we do not have a statistically complete sample, we should not only examine the statistical significances but also
the possible effect of biases.

We find that in all cases the best-fit values of the parameters and their 68\% credible intervals remain
naturally contained within the range set by the priors, and in most cases this is still true for the 95\% credible
interval.% , so our assumption of priors has little or no effect on the results. 
We also find that in individual cases there
can be strong correlations between parameter errors due to degeneracies in a specific fit, but we did not find any that 
were common enough to induce correlations between parameters in the overall population. 
We also find that for all physical 
parameters the range of best-fit values is significantly larger than the error regions of the better constrained
afterglows. This implies that there are no physical parameters, specifically also not the shock microphysics parameters or the beaming-corrected energy, that prefer a universal value. This is in agreement with previous studies (e.g., \citealt{2008ApJ...672..433S, 2009MNRAS.395..580C, 2015ApJ...799....3R}), but
now for a large and uniformly analysed sample of GRBs.

For our further description of the results, we focus on groups of physical parameters, from the outside in.
We begin with the ambient density, since this is the first distinction we make, and it is made in a way somewhat
different to the others, by comparing two different model fits. All others are simply free parameters fit within a
certain constrained but continuous range. Of these, we first discuss the energy and geometrical parameters
(opening angle and viewing angle), and after that the shock microphysics parameters. 

\subsection{GRB environment and ambient medium \label{sec:resenv}}

We do not assume a priori which model, homogeneous or wind-like environment, should be chosen for a given data set. Instead, we model every data set both for homogeneous and wind-like environment models, and choose which one explains the data best. Model selection is performed by comparing the evidence values from both fits. We present the log--evidence values, along with the corresponding Bayes factors, for each modelling effort in Table \ref{tab:evidence}. The Bayes factor, i.e. ratio of the evidence values, allows us to quantify the likelihood of the preferred model over the alternative model. A Bayes factor larger than 20 (e.g. \citealt{10.2307/2291091}) suggests a strong preference for the selected model. 15 out 26 GRBs in our sample, all long, have a Bayes factor larger than 20, and 8 out of these GRBs show evidence for a constant density environment. Thus, if we only consider the GRBs with a strong preference, there is an approximately even split between homogeneous and wind-like environments. \cite{2008ApJ...672..433S} have analyzed a sample of 10 GRBs and commented on their CBM density profile. They also find that both ISM-like and wind-like environments are required to explain the observed light curves for their sample of GRBs. \cite{2009MNRAS.395..580C} have analyzed the optical and X-ray light curves of 10 GRB afterglows, and arrived at the same conclusion. \cite{2011A&A...526A..23S} have compiled a sample of 27 \textit{Swift} detected GRBs (including one short GRB), and utilized the observed X-ray and optical afterglow emission to comment on the density profiles of their environments. They are able to determine that 18 GRBs in their sample are consistent with homogeneous environments, and 6 GRBs are consistent with wind-like environments. If we do not restrict ourselves to high Bayes factors, a slightly
higher fraction of afterglows favours an ISM solution (16 out of 25, i.e., 64\%).

For the short GRB sample the evidence values for both models are closer to each other. This is mainly due to the fact that short GRBs have fewer observations available, and therefore the data sets are less constraining; importantly, none of the short GRBs favour a wind environment, in agreement with the usual notion that they occur in less dense and near-uniform ISM. For short GRB 130603B, the evidence values for homogeneous and wind-like environments are consistent with each other considering the evidence uncertainty. Given that an ISM environment is a priori favoured, we chose that solution. 

The wider environment of the GRB is also probed by the host extinction, $A_V$. This is of course biased
to somewhat low values by the fact that we want well-detected optical afterglows, and for short GRBs to somewhat higher values because we need them to lie in regions of not too low density to produce a detectable afterglow. We find that more than half the afterglows have a nonzero $A_V$ with better than $2\sigma$ significance, with no significant differences between short and long GRBs or wind and uniform ambient media.

\begin{table*}
    \setlength{\tabcolsep}{5pt}
    \def\arraystretch{1.25}
    \caption{Inferred physical parameters for the GRB sample. The presented values correspond to the mode of the obtained posterior distribution. Reported uncertainties represent the 68\% credible interval . $k$ represents the CBM density profile (see Equation~\ref{eq:densityprofile}) and is either 0 for homogeneous or 2 for wind-like environments. }
    \begin{tabular}{|c|l|r|r|r|r|r|r|r|r|r}
    \hline
    \multicolumn{1}{l}{} & \multicolumn{1}{l}{Burst name} & \multicolumn{1}{r}{$\log_{10}\theta_0$ [rad]} & \multicolumn{1}{r}{$\log_{10}E_{K, \mathrm{iso}}$ [erg]} & \multicolumn{1}{r}{$\log_{10}n_\mathrm{ref}$} & \multicolumn{1}{r}{$\theta_{\mathrm{obs}} / \theta_0$} & \multicolumn{1}{r}{$p$} & \multicolumn{1}{r}{$\log_{10}\epsilon_B$} & \multicolumn{1}{r}{$\log_{10}\bar{\epsilon}_e$} & \multicolumn{1}{r}{$A_V$} & \multicolumn{1}{r}{$k$} \\
    \hline
    \hline
    \parbox[t]{2mm}{\multirow{4}{*}{\rotatebox[origin=c]{90}{short GRBs}}} & 051221A & $-0.98^{+0.13}_{-0.17}$ & $54.48^{+0.80}_{-1.69}$ & $0.55^{+1.26}_{-1.80}$ & $0.92^{+0.11}_{-0.29}$ & $1.78^{+0.17}_{-0.10}$ & $-2.40^{+0.91}_{-2.17}$ & $-5.26^{+2.43}_{-1.25}$ & $0.10^{+0.26}_{-0.10}$ &0\\
    &130603B & $-0.97^{+0.28}_{-0.54}$ & $53.89^{+0.82}_{-1.17}$ & $-1.14^{+0.59}_{-1.36}$ & $0.83^{+0.13}_{-0.36}$ & $2.67^{+0.25}_{-0.38}$ & $-4.71^{+1.72}_{-3.85}$ & $-0.98^{+0.51}_{-0.56}$ & $0.74^{+0.44}_{-0.32}$ &0\\
    &140903A & $-1.15^{+0.35}_{-0.22}$ & $54.58^{+0.57}_{-1.20}$ & $-1.97^{+1.17}_{-0.82}$ & $0.41^{+0.26}_{-0.33}$ & $2.36^{+0.17}_{-0.53}$ & $-7.21^{+3.58}_{-1.82}$ & $-1.07^{+0.63}_{-1.98}$ & $0.22^{+0.32}_{-0.22}$ &0\\
    &200522A & $-0.22^{+0.39}_{-0.31}$ & $53.54^{+1.08}_{-0.78}$ & $1.94^{+0.51}_{-1.48}$ & $0.51^{+0.35}_{-0.20}$ & $1.84^{+0.06}_{-0.06}$ & $-7.75^{+1.89}_{-1.27}$ & $-3.39^{+1.32}_{-1.33}$ & $5.29^{+2.79}_{-3.33}$ &0\\
    \hline
    \hline
    \parbox[t]{2mm}{\multirow{22}{*}{\rotatebox[origin=c]{90}{long GRBs}}} & 970508 & $0.06^{+0.03}_{-0.05}$ & $53.20^{+0.22}_{-0.21}$ & $2.18^{+0.11}_{-0.14}$ & $1.08^{+0.04}_{-0.03}$ & $2.57^{+0.05}_{-0.05}$ & $-4.39^{+0.38}_{-0.27}$ & $-0.49^{+0.05}_{-0.09}$ & $0.13^{+0.07}_{-0.05}$ &2\\
    &980703 & $-0.34^{+0.26}_{-0.20}$ & $52.28^{+0.36}_{-0.19}$ & $0.50^{+0.25}_{-0.30}$ & $0.89^{+0.16}_{-0.55}$ & $2.07^{+0.05}_{-0.10}$ & $-0.53^{+0.43}_{-0.57}$ & $-1.89^{+0.20}_{-0.17}$ & $1.01^{+0.16}_{-0.12}$ &2\\
    &990510 & $-1.16^{+0.03}_{-0.05}$ & $52.99^{+0.11}_{-0.08}$ & $-0.95^{+0.15}_{-0.35}$ & $0.31^{+0.13}_{-0.07}$ & $1.93^{+0.06}_{-0.05}$ & $-1.25^{+0.42}_{-0.45}$ & $-1.54^{+0.09}_{-0.12}$ & $0.01^{+0.02}_{-0.01}$ &0\\
    &991208 & $-1.94^{+0.19}_{-0.06}$ & $54.64^{+0.10}_{-0.41}$ & $-0.60^{+0.19}_{-0.11}$ & $1.05^{+0.41}_{-0.82}$ & $1.61^{+0.08}_{-0.06}$ & $-0.07^{+0.07}_{-0.16}$ & $-1.33^{+0.12}_{-0.11}$ & $0.44^{+0.09}_{-0.08}$ &0\\
    &991216 & $-0.59^{+0.37}_{-0.25}$ & $53.71^{+0.23}_{-0.44}$ & $1.28^{+0.23}_{-0.52}$ & $0.79^{+0.14}_{-0.41}$ & $2.14^{+0.20}_{-0.08}$ & $-3.61^{+0.54}_{-0.82}$ & $-1.27^{+0.23}_{-0.23}$ & $0.07^{+0.11}_{-0.06}$ &2\\
    &000301C & $-0.69^{+0.03}_{-0.03}$ & $52.42^{+0.16}_{-0.08}$ & $0.33^{+0.47}_{-0.11}$ & $0.07^{+0.04}_{-0.07}$ & $1.88^{+0.10}_{-0.06}$ & $-0.21^{+0.14}_{-0.78}$ & $-1.52^{+0.19}_{-0.07}$ & $0.03^{+0.08}_{-0.02}$ &0\\
    &000418 & $-0.26^{+0.27}_{-0.29}$ & $54.47^{+1.00}_{-0.82}$ & $1.11^{+0.70}_{-1.12}$ & $0.26^{+0.28}_{-0.20}$ & $2.38^{+0.15}_{-0.18}$ & $-1.91^{+0.52}_{-4.22}$ & $-1.59^{+0.45}_{-0.50}$ & $1.22^{+0.40}_{-0.29}$ &2\\
    &000926 & $-0.24^{+0.31}_{-0.28}$ & $55.13^{+0.63}_{-0.46}$ & $1.99^{+0.79}_{-0.39}$ & $0.34^{+0.20}_{-0.33}$ & $2.96^{+0.04}_{-0.04}$ & $-6.21^{+1.51}_{-1.08}$ & $-0.95^{+0.34}_{-0.28}$ & $0.20^{+0.05}_{-0.04}$ &2\\
    &010222 & $-0.40^{+0.05}_{-0.22}$ & $53.94^{+0.17}_{-0.15}$ & $-2.32^{+0.19}_{-0.17}$ & $0.13^{+0.29}_{-0.09}$ & $2.62^{+0.03}_{-0.04}$ & $-4.18^{+0.52}_{-0.31}$ & $-0.47^{+0.06}_{-0.08}$ & $0.54^{+0.05}_{-0.05}$ &0\\
    &030329 & $0.20^{+0.01}_{-0.07}$ & $53.04^{+0.10}_{-0.10}$ & $2.59^{+0.20}_{-0.18}$ & $0.73^{+0.05}_{-0.05}$ & $2.62^{+0.02}_{-0.06}$ & $-5.58^{+0.38}_{-0.33}$ & $-0.76^{+0.08}_{-0.06}$ & $0.01^{+0.02}_{-0.01}$ &0\\
    &050820A & $-0.44^{+0.24}_{-0.21}$ & $53.24^{+0.12}_{-0.11}$ & $0.95^{+0.23}_{-0.16}$ & $0.70^{+0.10}_{-0.39}$ & $2.11^{+0.14}_{-0.07}$ & $-1.95^{+0.19}_{-0.49}$ & $-1.31^{+0.25}_{-0.18}$ & $0.37^{+0.06}_{-0.08}$ &2\\
    &050904 & $-1.02^{+0.12}_{-0.04}$ & $53.31^{+0.28}_{-0.17}$ & $1.05^{+0.25}_{-0.87}$ & $0.54^{+0.21}_{-0.13}$ & $2.11^{+0.08}_{-0.08}$ & $-2.07^{+0.50}_{-0.46}$ & $-1.54^{+0.13}_{-0.40}$ & $0.07^{+0.05}_{-0.03}$ &0\\
    &060418 & $-0.82^{+0.06}_{-0.07}$ & $52.88^{+0.17}_{-0.14}$ & $0.23^{+0.56}_{-0.36}$ & $0.83^{+0.05}_{-0.06}$ & $2.28^{+0.06}_{-0.04}$ & $-2.81^{+0.32}_{-0.65}$ & $-1.30^{+0.17}_{-0.13}$ & $0.17^{+0.05}_{-0.05}$ &0\\
    &090328 & $-0.68^{+0.20}_{-0.13}$ & $53.17^{+0.51}_{-0.86}$ & $1.66^{+0.70}_{-0.79}$ & $0.94^{+0.23}_{-0.33}$ & $2.28^{+0.10}_{-0.18}$ & $-3.90^{+1.31}_{-0.89}$ & $-1.39^{+0.48}_{-0.37}$ & $0.07^{+0.11}_{-0.06}$ &0\\
    &090423 & $-0.06^{+0.26}_{-0.10}$ & $53.46^{+0.50}_{-0.48}$ & $2.01^{+0.75}_{-0.62}$ & $0.84^{+0.11}_{-0.46}$ & $2.16^{+0.17}_{-0.24}$ & $-4.81^{+1.48}_{-0.97}$ & $-1.41^{+0.35}_{-0.56}$ & $0.09^{+0.09}_{-0.07}$ &0\\
    &090902B & $0.04^{+0.11}_{-0.34}$ & $53.43^{+0.23}_{-0.37}$ & $0.50^{+0.58}_{-0.31}$ & $0.24^{+0.19}_{-0.22}$ & $2.23^{+0.04}_{-0.09}$ & $-3.15^{+0.70}_{-0.91}$ & $-1.57^{+0.29}_{-0.23}$ & $0.07^{+0.06}_{-0.06}$ &2\\
    &090926A & $-0.81^{+0.08}_{-0.10}$ & $53.09^{+1.17}_{-0.32}$ & $0.19^{+2.07}_{-0.26}$ & $0.78^{+0.25}_{-0.14}$ & $2.24^{+0.08}_{-0.05}$ & $-1.57^{+0.41}_{-2.14}$ & $-0.93^{+0.28}_{-0.81}$ & $0.09^{+0.01}_{-0.01}$ &0\\
    &120521C & $-0.91^{+0.11}_{-0.11}$ & $52.98^{+0.23}_{-0.16}$ & $-0.47^{+0.41}_{-0.29}$ & $0.50^{+0.20}_{-0.34}$ & $2.88^{+0.12}_{-0.25}$ & $-1.10^{+0.37}_{-0.51}$ & $-0.84^{+0.16}_{-0.16}$ & $0.78^{+0.09}_{-0.09}$ &0\\
    &130427A & $-0.23^{+0.25}_{-0.15}$ & $52.58^{+0.64}_{-0.17}$ & $0.17^{+0.33}_{-0.53}$ & $0.20^{+0.34}_{-0.20}$ & $2.16^{+0.04}_{-0.05}$ & $-2.31^{+0.62}_{-0.94}$ & $-1.13^{+0.20}_{-0.25}$ & $0.03^{+0.03}_{-0.03}$&2\\
    &130702A & $-0.71^{+0.12}_{-0.05}$ & $53.07^{+0.90}_{-0.97}$ & $-0.90^{+1.42}_{-0.45}$ & $0.04^{+0.11}_{-0.04}$ & $1.46^{+0.08}_{-0.04}$ & $-0.71^{+0.63}_{-1.24}$ & $-7.65^{+2.05}_{-1.89}$ & $0.26^{+0.18}_{-0.12}$ &0\\
    &130907A & $-1.38^{+0.17}_{-0.05}$ & $53.02^{+0.34}_{-0.09}$ & $-1.26^{+0.12}_{-0.11}$ & $0.88^{+0.23}_{-0.21}$ & $1.87^{+0.09}_{-0.09}$ & $-0.04^{+0.04}_{-0.10}$ & $-1.03^{+0.14}_{-0.18}$ & $1.60^{+0.15}_{-0.16}$ &0\\
    &140304A & $-1.30^{+0.23}_{-0.07}$ & $54.46^{+0.28}_{-0.42}$ & $1.96^{+0.57}_{-0.35}$ & $1.19^{+0.34}_{-0.13}$ & $2.28^{+0.17}_{-0.25}$ & $-3.73^{+0.76}_{-0.72}$ & $-1.40^{+0.20}_{-0.33}$ & $0.41^{+0.11}_{-0.08}$ &2\\
    \hline
    \hline
    \end{tabular}
    \label{tab:parameters}
\end{table*}

Now that we have found the best ambient-density model for each afterglow, we will look at the other parameters, for which we take the values for the best-fit ambient medium in each case.
We present our modelling results for the GRB parameters (including the rest-frame host extinction values, $A_V$) in Table \ref{tab:parameters}. In Figures \ref{fig:plt-parameters-ism}, \ref{fig:plt-parameters-wind} and \ref{fig:plt-parameters-sgrb} we present the posterior distribution for each parameter (in the form of a violin plot) together with their 68\% credible intervals. The complete set of light curves and posterior distributions are available as online supplementary material, including results for both homogeneous and wind-like environments for each GRB in our sample.

\subsection{Energy, opening angle, and viewing angle\label{sec:resene}}

\begin{figure*}
 \includegraphics[width=\textwidth]{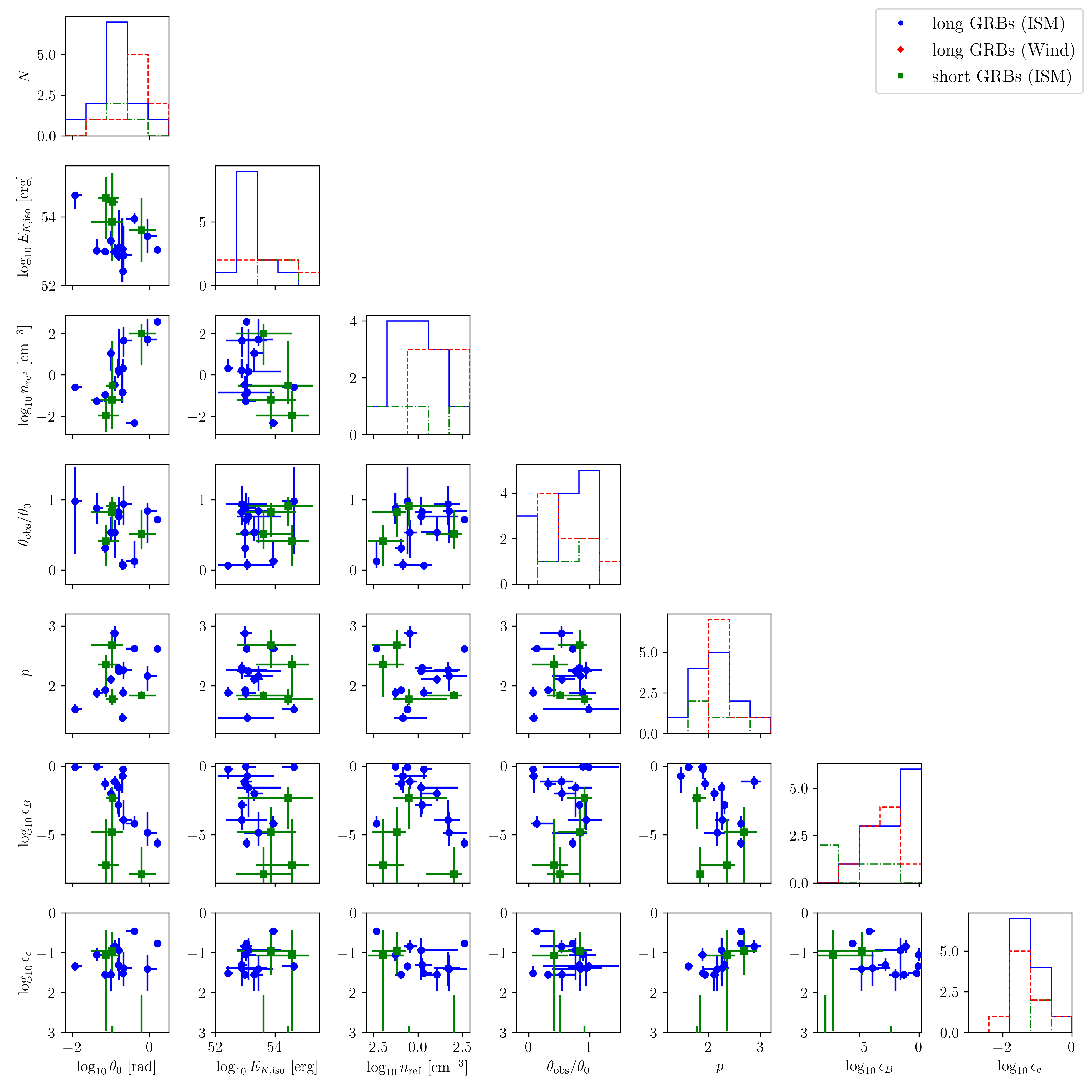}
 \caption{Corner plot for the inferred physical parameters of the GRBs associated with constant density environments. Blue circles and green squares represent the inferred parameter value of long GRBs and short GRBs, respectively. The error bars represent the 68\% credible limit. }
 \label{fig:plt-corner-ism}
\end{figure*}

\begin{figure*}
 \includegraphics[width=\textwidth]{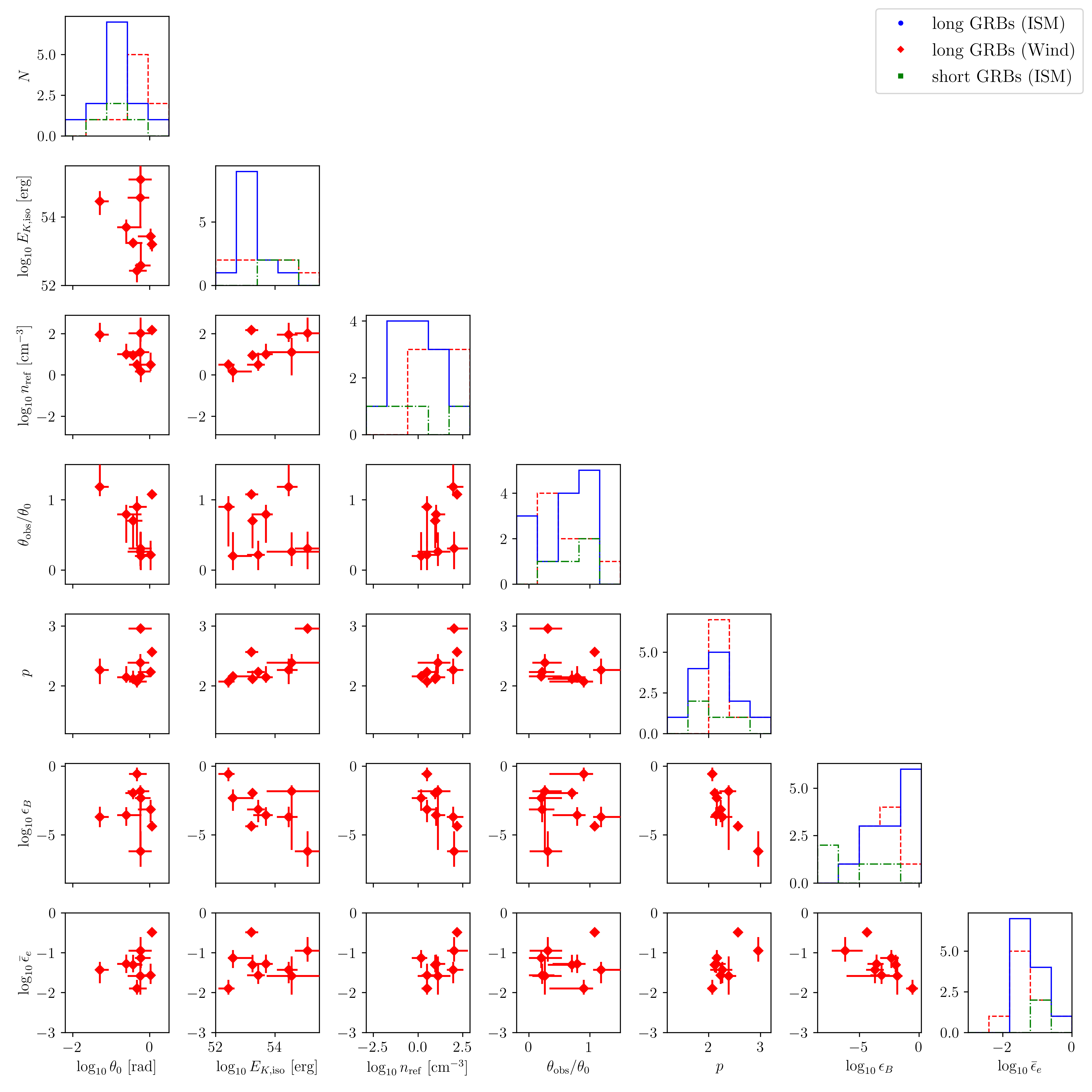}
 \caption{Corner plot for the inferred physical parameters of the GRBs associated with wind-like environments. The error bars represent the 68\% credible limit.}
 \label{fig:plt-corner-wind}
\end{figure*}

In Figures \ref{fig:plt-corner-ism} and \ref{fig:plt-corner-wind} we present the parameter values for the GRBs associated with homogeneous and wind-like environments, respectively, in the form of a corner plot. These figures help us to identify any correlations between the burst parameters. The diagonal elements in each figure contain the parameter distributions for the single fit parameters, with different colours for the short (green), long-ISM (blue) and long-wind (red) GRBs.

\textbf{Opening angle:} We do not find a notably different opening angle distribution for short and long GRBs. When a Kolmogorov-Smirnov (KS) test is performed on the inferred opening angle of ISM-like long GRBs and short GRBs, we find a $p$ value of 0.48 for the hypothesis that the two samples are drawn from the same distribution. 
A KS test checking the consistency of the $\theta_0$ distribution between ISM- and wind-like long GRBs yields we find $p=0.012$. On its own that might be considered moderate evidence for a difference, but given the many trials (distribution comparisons) in this paper, it is not (see above). 

\begin{figure}
 \includegraphics[width=\columnwidth]{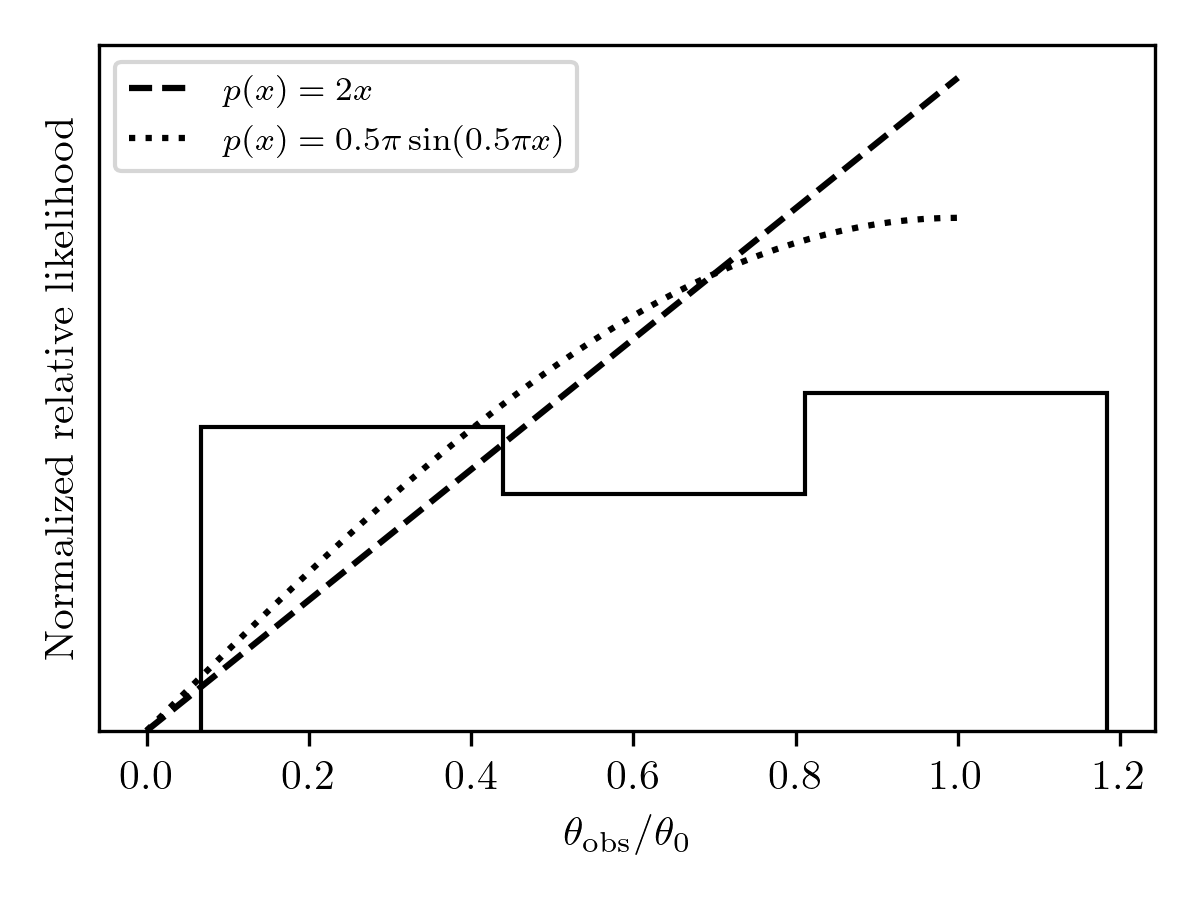}
 \caption{Histogram of the inferred $\theta_{\mathrm{obs}} / \theta_0$ for the GRB sample. The dotted line represents the analytically expected probability density function for large opening angles ($\theta_0 = \pi / 2\ \mathrm{rad}$), whereas the dashed line represents the probability density function for small opening angles ($\theta_0 \ll \pi / 2\ \mathrm{rad}$).}
 \label{fig:plt-theta_obs_frac}
\end{figure}

\textbf{Observer viewing angle:} The distribution of the observer viewing angle, $\theta_{\mathrm{obs}}$, does not follow a simple form, since it is constrained to be within the jet opening angle (at least at early times),  but that is 
different for each GRB as we have just seen. However, the fractional observer angle distribution, $\theta_{\mathrm{obs}} / \theta_0$, does have a simpler form under the top-hat jet assumption, because in that case every direction within the opening angle has the same properties and thus the same brightness at early times: its probability density is linear for $\theta_0 \ll \pi / 2$ and 
a sine function for $\theta_0=\pi / 2$. 
%A more centrally concentrated jet (Gaussian or power-law), in which the jet axis is brighter than the edge, should show a preference for small values of $\theta_{\mathrm{obs}} / \theta_0$. 
We show the distribution for the full sample in Figure \ref{fig:plt-theta_obs_frac},
with the two limiting theoretical cases. The observed distribution extends a bit beyond 1, but no values are significantly larger than 1. Even so, KS-comparison with the theoretical distributions gives $p=0.17$ for accepting the null hypothesis of equality. We conclude that the data are consistent with the top-hat jet hypothesis and do not strongly indicate a structured, more centrally concentrated jet (which would closely resemble a top-hat jet for observers close to the jet axis in any case, e.g., \citealt{2002ApJ...564..209D, 2002MNRAS.332..945R,2004MNRAS.354...86R,2003ApJ...591.1086G,2003ApJ...591.1075K,2003ApJ...592..390P,2003ApJ...592.1002S,2020ApJ...896..166R}). Specifically, our viewing angle has to be within the opening angle of the gamma-ray emission, which typically comes from material with $\Gamma>100$; If the early afterglow emission, of which we derive the opening angle in our fits, had a significantly greater opening angle, then we would find the distribution of $\theta_{\mathrm{obs}} / \theta_0$ biased towards small values. Since the afterglow represents all the material with $\Gamma\gtrsim10$, this means our result argues against jet structures in which the material with initial Lorentz factors above 10 has a significantly wider opening angle than the material with initial Lorentz factors above 100.
%However, this does argue against the opening angle of the prompt $\gamma$-ray emission %being significantly narrower than that of the afterglow emission, since all our GRBs are %gamma-ray selected, and thus a wider afterglow opening angle would lead to small values %of $\theta_\mathrm{obs}/\theta_0$. 
This may argue against, or at least significantly constrain so-called `jet-cocoon' models of GRBs, in which a core jet with quite high initial Lorentz factors ($\Gamma_0\gtrsim100$) is surrounded by an energetic cocoon with Lorentz factors of several tens (e.g., \citealt{2002MNRAS.337.1349R, 2005ApJ...626..966P}).
%Note that the simulations underlying our fitted models do include mass entrainment and thus the possibility of cocoon formation, but only in the ambient medium. 
The initial conditions of the simulations underlying our model are already outside the progenitor object. We do not think our afterglow selection has biased
us against jet-cocoon cases: if the cocoon only decelerated after our
afterglow data start, it would give rise to a late-injection or plateau phase in
the afterglow \citep{2006MNRAS.366L..13G, 2014MNRAS.445.2414V}, and we have not excluded any afterglows for obvious signs thereof. If the cocoon decelerated before our afterglow data start, then the tendency to smaller observer angles because of the gamma-ray selection would remain, and we do not see this.

The results on the GRB energy are more complex, and we defer them to the discussion section.

\begin{figure}
 \includegraphics[width=\columnwidth]{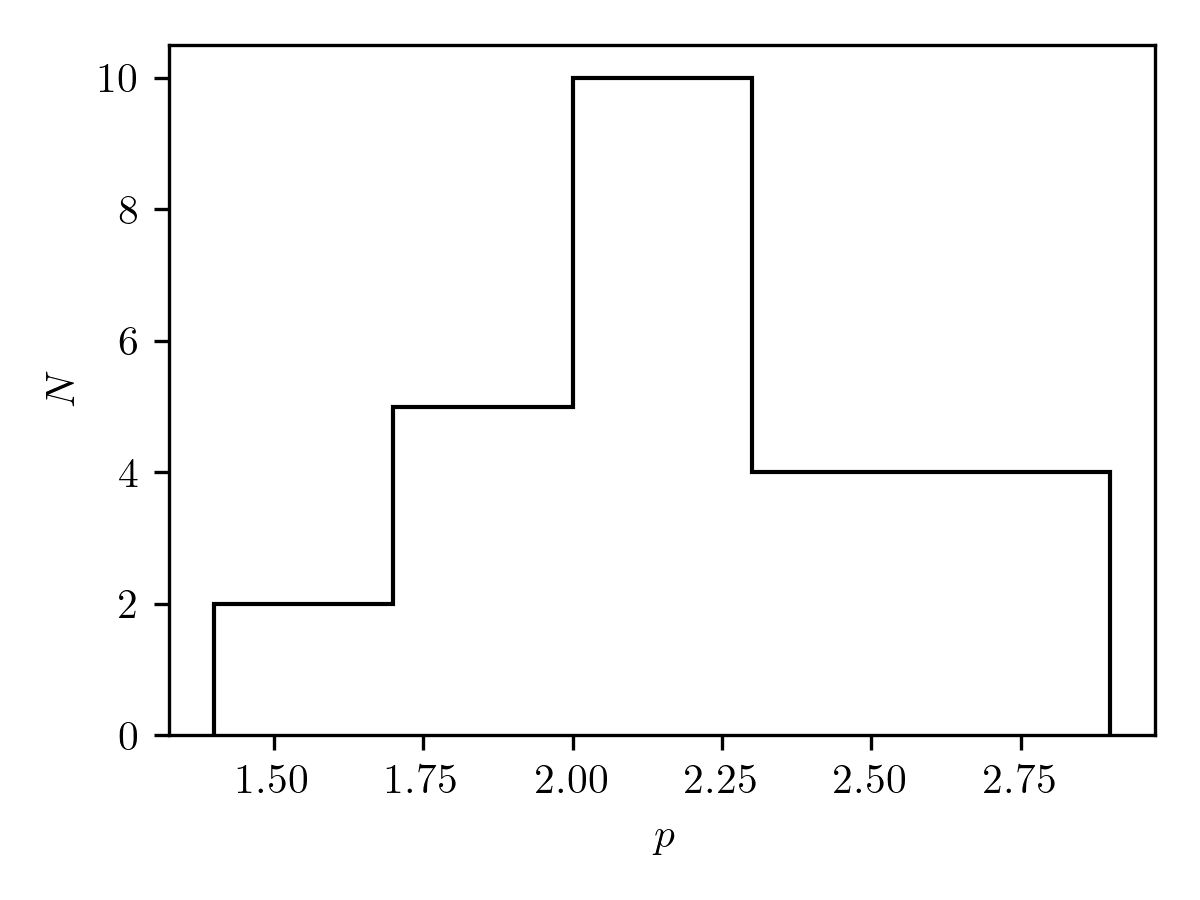}
 \caption{Histogram of the inferred $p$ for the GRB sample.}
 \label{fig:plt-hist-p}
\end{figure}

\subsection{Shock physics parameters\label{sec:ressho}}

For the shock physics parameters we have to be a bit careful (see Section~\ref{sec:model}): in order to avoid too strong
degeneracies for $p\simeq2$, we fit for $p$ and $\Bar{\epsilon}_e$, and indeed we do find some cases of $p<2$.
The distribution of $p$, the power-law index of the shock-accelerated particles, can be seen in Figure~\ref{fig:plt-hist-p}. We find that the $p$-values are consistent with being drawn from the same distribution for long GRBs and short GRBs. We find a mean value of $2.21$ and standard deviation $\sigma_p=0.36$ for the inferred $p$ values. \cite{2010ApJ...716L.135C} have analyzed a large sample of \textit{Swift} detected GRBs to determine the distribution of $p$. They utilized the reported spectral indices in X rays to determine the $p$ values of their sample, using closure relations. They find that the distribution of $p$ is consistent with a Gaussian distribution with $\mu=2.36$ and $\sigma=0.59$; given the errors in both methods, we consider the two results to be consistent. We find three cases where $p$ is significantly less than 2, and thus where a high-energy cutoff to the electron distribution is required to keep the
total electron energy finite. In more than half the cases (15), $p=2$ is included within the 95\% confidence 
region of the fit result, implying that indeed using $\Bar{\epsilon}_e$ is required to avoid problems in the fitting process.

While the short GRBs all have  $\epsilon_B$ values on the low side, their small number and large error bars prevent us from drawing any strong conclusions in this respect: the KS test results in $p=0.10$ for the short/long GRB distributions of magnetic-field energy fractions being the same, and similarly we find no evidence for a difference between the
long GRBs in ISM and wind environments. What is quite striking though is that $\epsilon_B$ ranges over 5--6 decades in value, a much greater range than $\epsilon_e$.

\begin{figure}
 \includegraphics[width=\columnwidth]{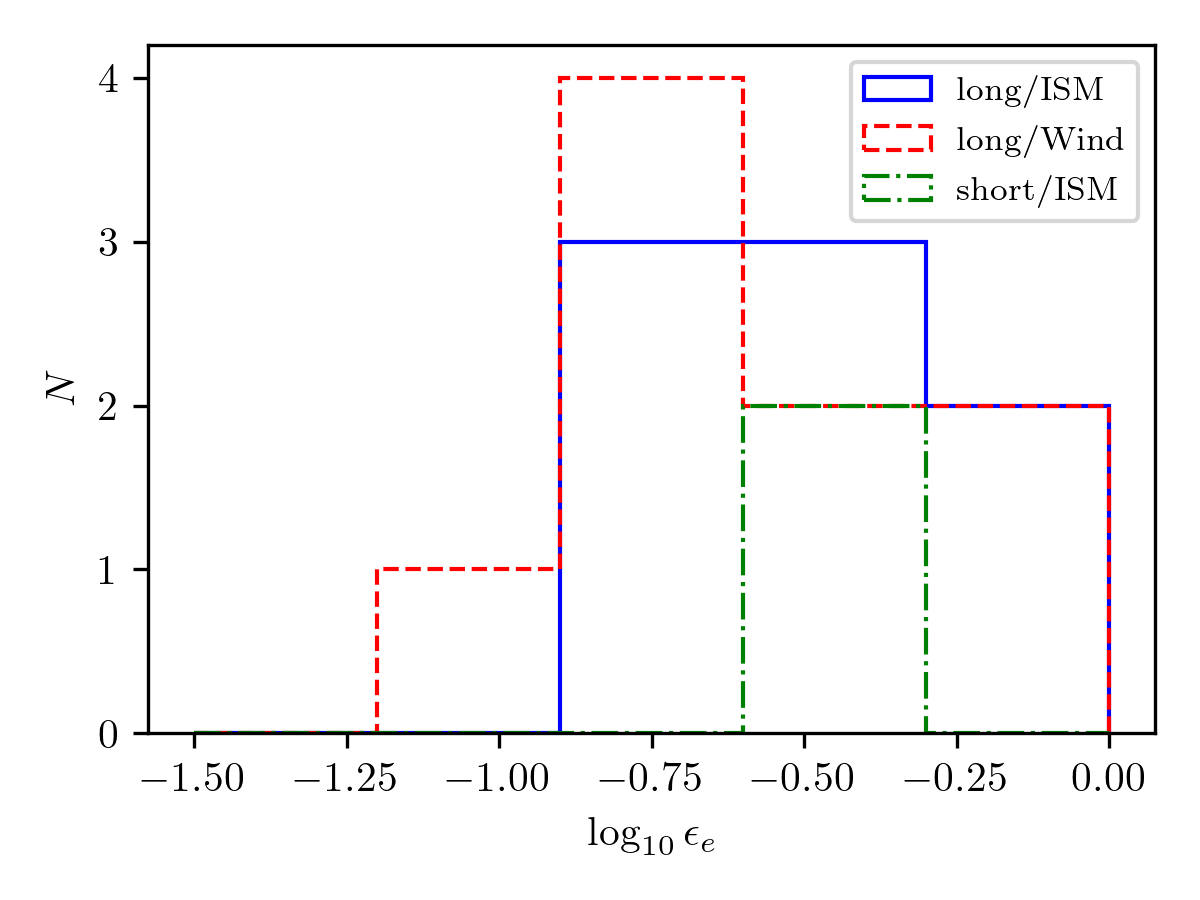}
 \caption{Histogram of the inferred $\epsilon_e$ for the GRB sample.}
 \label{fig:plt-hist-epsilon_e}
\end{figure}

Since $\epsilon_e$ is a physically more meaningful measure of the electron energy density, we derive it from
the nominal fit values in case $p>2$. The derived $\epsilon_e$ values can be seen in Table~\ref{tab:derived_parameters}. In Figure~\ref{fig:plt-hist-epsilon_e} we present the $\epsilon_e$ distribution of the GRB sample, only for GRBs with inferred mode value of $p>2$. We find that $\epsilon_e$
is never very low and always above 0.1, with some values (uncomfortably?) close to 1. The values for
the different subsamples are in good agreement (mean values are 0.34 for homogeneous environment and 0.28
for wind).
\cite{2017MNRAS.472.3161B} have demonstrated that it is possible to constrain $\epsilon_e$ by measuring the peak flux and peak time of the radio afterglow light curve. By applying this method to a sample of 36 long GRBs, they were able to put upper limits on the scatter of $\epsilon_e$. They find that $\sigma_{\log_{10}\epsilon_e} < 0.31$ for constant density environments, and $\sigma_{\log_{10}\epsilon_e} < 0.26$ for wind-like environments. We find that the standard deviation of $\epsilon_e$ for the long GRB sample is $\sigma_{\log_{10}\epsilon_e} = 0.24$ for homogeneous environments and $\sigma_{\log_{10}\epsilon_e} = 0.28$ for wind-like environments. Note that, although the standard deviations of the inferred $\epsilon_e$ distributions are consistent with \cite{2017MNRAS.472.3161B}, they find lower mean values of 0.15 and 0.13 for ISM-like and wind-like long GRBs, respectively.

\section{Discussion}
\label{sec:discussion}

\subsection{GRB environment and ambient medium\label{sec:discenv}}

The inferred CBM densities for ISM-like long and short GRBs exhibit a wide distribution. The mean value for the circumburst densities of ISM-like long and short GRBs are $1.26$ and $0.39$ $\mathrm{cm^{-3}}$ with standard log-deviations of $\sigma_{\log_{10}n_{\mathrm{ref}}} = (1.32, 1.49)$, respectively, i.e., they cover about 3 decades in density.
Given the wide variety of possible massive-star and merger environments, this is not so surprising.
 
We do not find any pronounced differences between the density distributions for short GRBs and ISM-like long GRBs. Canonically, it is expected that short GRB progenitors should be in lower-density environments than long GRBs. \cite{2013ApJ...776...18F} report that short GRBs are localized to lie at greater distances from their host galaxy centres when compared to long GRBs. They find that, for short GRBs, the median value of the offset from their galaxian centre is 4.5\,kpc, and when compared to the size of their host galaxy the median value of the offset becomes $r / r_{\mathrm{host}} = 1.5$. Note that there is a strong bias in our short-GRB sample because we require bright afterglows, and afterglow brightness goes up strongly with ambient density.
And indeed, if we check our four short GRBs we find that their environment is quite atypical for the short-GRB population: studies of the host galaxies of our four short GRBs show that they do not have a large offset from their host centre. GRB 051221A has been identified to lie in a star-forming galaxy with an estimated normalized offset of $r / r_{\mathrm{host}} = 0.29 \pm 0.04$ \citep{2006ApJ...650..261S}. GRB 130603B is associated with a spiral galaxy, and has been localized to a tidally disrupted arm at a distance of $5.4 \pm 0.3$\,kpc from the centre of the galaxy \citep{2014A&A...563A..62D}. \cite{2016ApJ...827..102T} find that GRB 140903A lies at a distance of $0.5 \pm 0.2$\,kpc from the centre of its host. \cite{2021ApJ...906..127F} estimate a normalized offset of $r / r_{\mathrm{host}} = 0.24 \pm 0.04$ for GRB 200522A (see also \citealt{2021MNRAS.502.1279O}). This is quite unlike the full population: \cite{2015ApJ...815..102F} have studied the afterglow emission from a sample of 38 short GRBs and have found that they lie in low-density environments with median densities of $(3\mathrm{-}15)\times10^{-3}\ \mathrm{cm^{-3}}$. They also state that 80 to 95\% of short GRBs in their sample have densities smaller than $1\ \mathrm{cm^{-3}}$, which is also true for 3 out of 4 GRBs in our short GRB sample within the reported uncertainties. \cite{2020MNRAS.495.4782O} have utilized the X-ray light curves of the \textit{Swift} population of short GRBs to constrain their circumburst environment densities. They assumed fiducial values for the GRB parameters and have found that $\lesssim16\%$ of the population have densities lower than $10^{-4}\ \mathrm{cm^{-3}}$, and that $\gtrsim30\%$ of the population has densities larger than $10^{-2}\ \mathrm{cm^{-3}}$. In other words, our requirement of a well-detected and well-sampled afterglow does seem to have biased our short GRBs to lie in regions similar to those of long GRBs. This means they are not very representative of the whole population of short GRBs. There may be a silver lining to this cloud, in that when discussing their physical parameters in comparison to long GRBs, we can eliminate strong fit-induced correlations with environmental 
parameters as a potential cause for any differences we find. Also, under the most likely scenario that short GRBs are all mergers, born from a binary long before the merger time, there is no reason to think that the physical properties of the merger and GRB explosion would depend on the medium they happen to be in at the time of merger. Hence, we do not think that this environmental bias makes our GRB sample biased relative to the whole short-GRB population in intrinsic parameters.
  
It would be good again to caution, now quantitatively, that for $k=0$, the radius of the blast wave scales with observer time as $r\propto t^{1/4}$, which means observations do not cover a wide range of radii. The afterglow starts at the deceleration radius, $r_\mathrm{dec}$, where half the initial jet energy has been deposited, and transitions into a spherical supernova remnant-like evolution at the non-relativistic radius $r_\mathrm{NR}$. For typical values these
are just under $10^{17}$\,cm and $10^{18}$\,cm, respectively, both scaling as $(E/n)^{1/3}$. Their ratio is
$r_\mathrm{NR}/n_\mathrm{dec}=20 (\Gamma_0/100)^{-2/3}$, where $\Gamma_0$ is the initial jet Lorentz factor. 
Given that we have few examples where we get close to either end of these regimes, we see that indeed the typical uniform-like afterglow covers only a small range of radii. Therefore, approximate uniformity of the ambient medium is enough, which may apply to many plausible environments. 

For long GRBs in wind-like environments, the distribution of $n_{\mathrm{ref}}$ has a mean of $14$ $\mathrm{cm^{-3}}$, or a mean $A$ value of $0.48$ $A_*$, with a standard log-deviation of 
$\sigma_{\log_{10}n_{\mathrm{ref}}} = 0.69$, rather narrower than the density range of the total sample. 
This indicates that the free-wind parameters we find are indeed similar to canonical values expected of massive
Wolf-Rayet stars, the most likely progenitors. It might argue that the likeliest reason for seeing uniform media
about equally often is that the reverse shock in the stellar-wind bubble is close enough that in many cases the
main afterglow phase is in the shocked wind. For the free-wind case, $r\propto t^{1/2}$, so the afterglow samples a markedly larger range of radii. For canonical values ($\dot{M}=10^{-5}\,\mathrm{M}_\odot$/yr, $E=10^{52}$\,erg,
$\Gamma_0=100$) we find $r_\mathrm{dec}=1.5\times10^{11}$\,cm and $r_\mathrm{NR}=1.5\times10^{15}$\,cm,
and since they again scale the same with most parameters, $r_\mathrm{NR}/r_\mathrm{dec}=10^4(\Gamma_0/100)^2$,
a large range indeed even if we see only part of it sampled in the data. Note that for a canonical
wind velocity of 1000\,km/s, the afterglow phase starts in the wind material that was emitted \textit{less than one hour before the star explodes}, and even in weak winds typically still less than one day. It ends in wind material that was
emitted a fraction of a year to a few years before the explosion. GRB afterglows thus do not probe
typical mean-life stellar wind parameters, and indicate that even very close to the end of the star's life the wind is
similar to that during an average moment in its life. 
We note that some, especially single-star, GRB scenarios prefer low mass loss rates of the progenitors, and it is known that in low-metallicity
environments massive stars do indeed have lower mass loss rates \citep{Vink+2001,Vink+2005};
since we find that $A/A_*=0.48$ on average, our fits do not provide evidence for this. This may agree will with more recent findings that GRBs actually
do not prefer low-metallicity environments, other than that they are suppressed in regions with metallicity above solar \citep[][, and references therein]{Perley+2016,Fynbo+2009}. This in turn may favour binary evolution scenarios for the origin of long GRBs \citep{Perley+2016, 2020MNRAS.491.3479C}.

These ambient-medium considerations ask for better investigation of scenarios in which the
blast wave emerges from the free wind early, since in light of the above this appears difficult, and yet we
find half or more of the afterglow fits prefer a uniform-medium solution.

\subsection{Energy and opening angle\label{sec:discene}}

The observed flux of GRB afterglow emission does not directly depend on the true energy of the burst, but rather depends directly on the energy per unit solid angle, or the isotropic equivalent energy, $E_\mathrm{iso}$. Therefore, to measure the true energetics of these events by afterglow modelling, we need to constrain both the isotropic equivalent energy and the opening angle. The true energy can then be calculated using
\begin{equation}
    E_{\mathrm{true}} = E_{\mathrm{iso}} (1 - \cos{\theta_0}).
    \label{eq:true-energy}
\end{equation}
This equation is valid for both the total energy and for the $\gamma$-ray and afterglow kinetic energies
separately, provided that the opening of both is the same; we have argued above (Sect.~\ref{sec:resene}) that this is indeed the case.

\begin{figure}
 \includegraphics[width=\columnwidth]{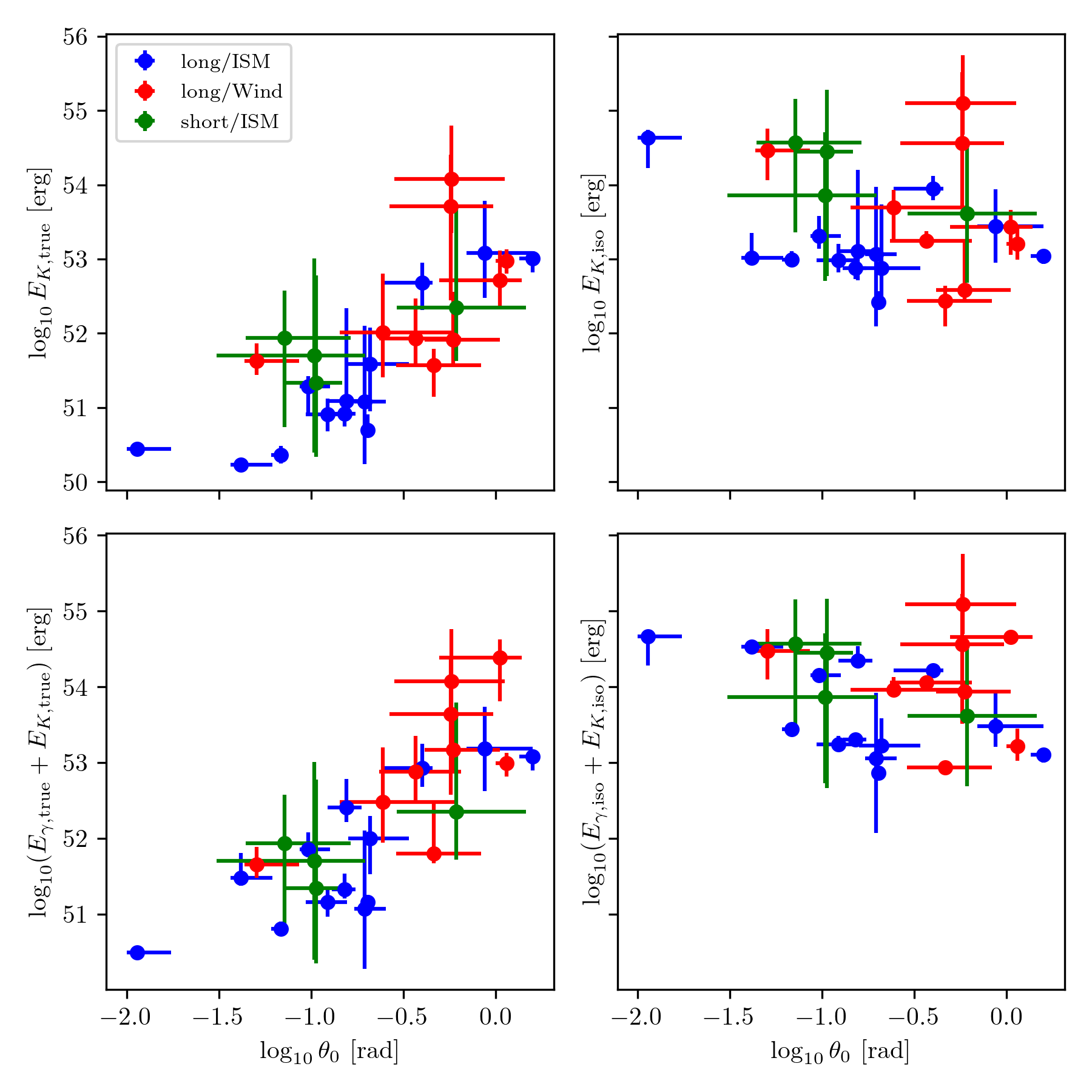}
 \caption{Scatter plot of $(E_{K, \mathrm{iso}}, E_{K, \mathrm{true}})$ and $(E_{total, \mathrm{iso}}, E_{total, \mathrm{true}})$ w.r.t. the inferred opening angles for the GRB sample. The blue circles and red diamonds represent long GRBs in homogeneous and wind-like environments, respectively. Error bars represent the 68\% credible interval.}
 \label{fig:plt-corner-theta_0-E_K_iso}
\end{figure}

\begin{figure}
 \includegraphics[width=\columnwidth]{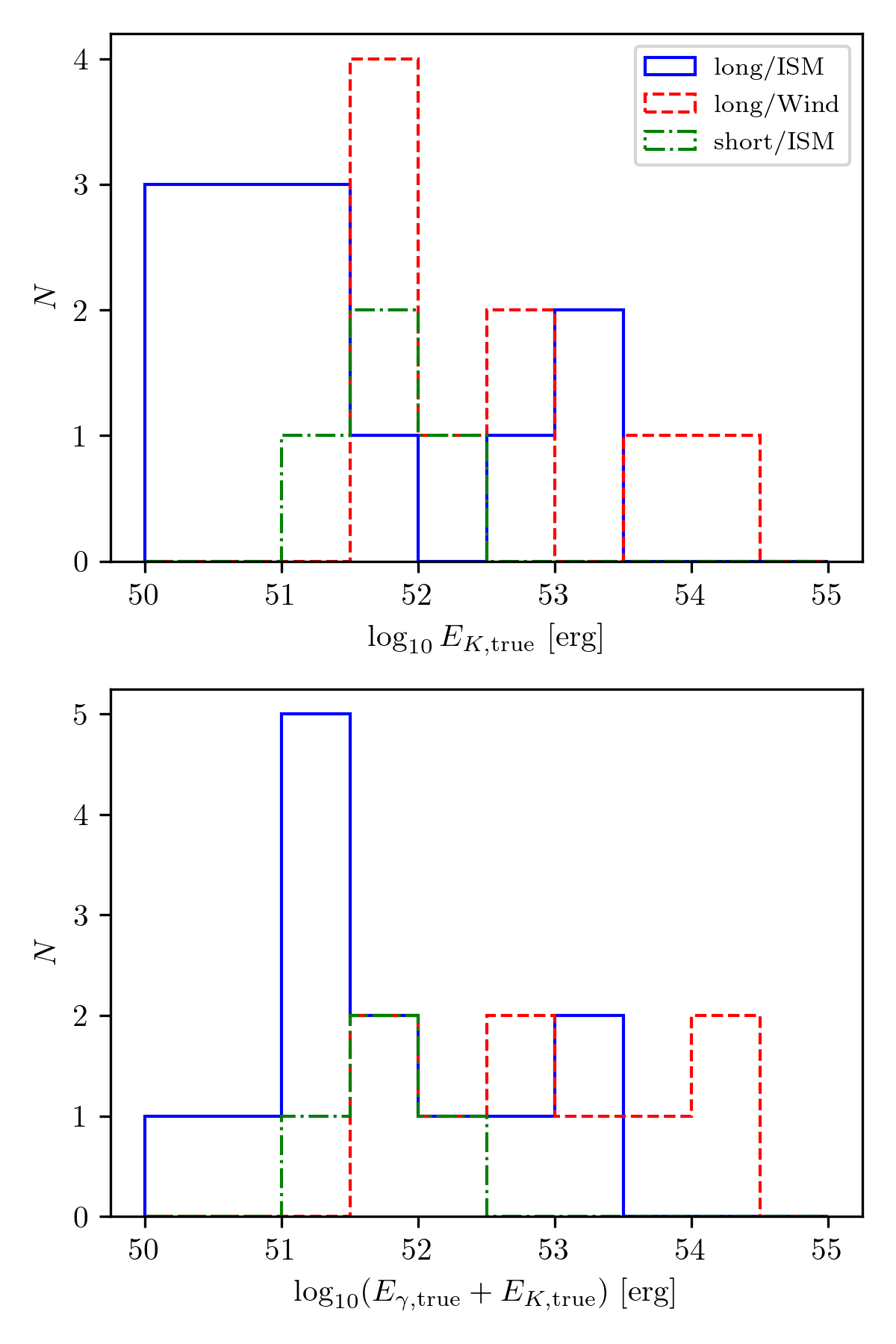}
 \caption{Histogram of the inferred energetics for the GRB sample. The upper and lower panels show the histograms for the $E_{K, \mathrm{true}}$ and $E_{\mathrm{true}}$ distributions, respectively.}
 \label{fig:E}
\end{figure}

In A20, we suggested the existence of a common kinetic energy reservoir for long GRBs, based on a very
small sample of only 5 GRBs. For the larger sample in this work, where we have also allowed both wind and ISM
ambient media, this no longer holds: in Figure~\ref{fig:plt-corner-theta_0-E_K_iso}, we present scatter plots of the isotropic-equivalent and beaming corrected kinetic/total energies with respect to the opening angles of our GRB sample. We perform a KS test on the mode of the posterior distribution for each GRB to determine whether or not the $E_{K, \mathrm{true}}$ values for homogeneous and wind-like GRBs are drawn from the same distribution. We find that we can reject the null hypothesis with a p-value of $6.5 \times 10^{-3}$,
which according to our strict criteria is not significant enough. The histograms of the true kinetic and total energies of the three sub-samples are shown in Figure~\ref{fig:E}. While these again have the disadvantage of making use only of the mode of the posterior distribution rather than all the information in the posterior distribution, they do illustrate some difference between the wind- and ISM-like long GRBs.

The $E_{K, \mathrm{true}}$ distributions for long and short GRBs are consistent with each other. This suggests that the kinetic energy of the explosion is not significantly different between short and long GRBs. However, the measured prompt emission energies are orders of magnitudes lower for short GRBs, which implies that it is the prompt emission efficiency of short GRBs that is lower. The prompt emission efficiency can be defined as
\begin{equation}
    \epsilon_\gamma \equiv \frac{E_{\gamma, \mathrm{true}}}{E_{\gamma, \mathrm{true}} + E_{K, \mathrm{true}}}.
\end{equation}

\begin{table*}
    \setlength{\tabcolsep}{5pt}
    \def\arraystretch{1.25}
    \caption{Derived parameters for the GRB sample. The presented values correspond to the mode of the obtained posterior distribution. Reported uncertainties represent the 68\% credible interval. We calculate $\epsilon_e$ values only for GRBs for which the inferred mode of $p$ is larger than 2. Missing values are represented by -. $k$ represents the CBM density profile (see Equation~\ref{eq:densityprofile}) and is either 0 for homogeneous or 2 for wind-like environments.}
    \label{tab:derived_parameters}
    \begin{tabular} {|c|l|r|r|r|r|r|r|}
        \hline
        \multicolumn{2}{c}{Burst name} &  \multicolumn{1}{c}{$\log_{10}\epsilon_e$} &  \multicolumn{1}{c}{$\log_{10}E_{K, \mathrm{true}}$ [erg]} &
        \multicolumn{1}{c}{$\log_{10}E_{\gamma, \mathrm{true}}$ [erg]} & \multicolumn{1}{c}{$\log_{10}E_{\mathrm{true}}$ [erg]} & \multicolumn{1}{c}{$\log_{10}\epsilon_\gamma$} & $k$ \\ 
        \hline
        \hline
        \parbox[t]{2mm}{\multirow{4}{*}{\rotatebox[origin=c]{90}{short GRBs}}} & 051221A & - & $51.33^{+1.45}_{-1.00}$ & $48.93^{+0.28}_{-0.35}$ & $51.34^{+1.44}_{-0.99}$ & $-3.27^{+1.78}_{-0.71}$ & $0$ \\
        & 130603B & $-0.44^{+0.35}_{-0.39}$ & $51.70^{+1.31}_{-1.31}$ & $49.05^{+0.58}_{-1.03}$ & $51.70^{+1.31}_{-1.30}$ & $-2.55^{+1.15}_{-0.84}$ & $0$ \\
        & 140903A & $-0.41^{+0.29}_{-0.31}$ & $51.94^{+0.64}_{-1.20}$ & $47.19^{+0.72}_{-0.42}$ & $51.94^{+0.64}_{-1.20}$ & $-4.79^{+1.21}_{-0.58}$ & $0$ \\
        & 200522A & - & $52.35^{+1.36}_{-0.72}$ & $49.19^{+0.71}_{-0.61}$ & $52.35^{+1.45}_{-0.63}$ & $-3.69^{+0.93}_{-0.95}$ & $0$ \\
        \hline
        \hline
        \parbox[t]{2mm}{\multirow{22}{*}{\rotatebox[origin=c]{90}{long GRBs}}} & 970508 & $-0.04^{+0.04}_{-0.08}$ & $52.98^{+0.15}_{-0.18}$ & $51.55^{+0.06}_{-0.10}$ & $52.99^{+0.14}_{-0.18}$ & $-1.43^{+0.20}_{-0.23}$ & $2$ \\
        & 980703 & $-0.68^{+0.23}_{-0.29}$ & $51.57^{+0.22}_{-0.43}$ & $51.86^{+0.50}_{-0.40}$ & $51.80^{+0.65}_{-0.13}$ & $-0.10^{+0.04}_{-0.09}$ & $2$ \\
        & 990510 & - & $50.36^{+0.12}_{-0.12}$ & $50.62^{+0.07}_{-0.10}$ & $50.81^{+0.09}_{-0.09}$ & $-0.19^{+0.03}_{-0.04}$ & $0$ \\
        & 991208 & - & $50.44^{+0.05}_{-0.04}$ & $49.16^{+0.37}_{-0.11}$ & $50.49^{+0.04}_{-0.04}$ & $-1.32^{+0.38}_{-0.09}$ & $0$ \\
        & 991216 & $-0.52^{+0.29}_{-0.16}$ & $52.02^{+0.79}_{-0.61}$ & $52.30^{+0.78}_{-0.47}$ & $52.48^{+0.72}_{-0.54}$ & $-0.13^{+0.07}_{-0.17}$ & $2$ \\
        & 000301C & - & $50.70^{+0.21}_{-0.08}$ & $50.97^{+0.07}_{-0.05}$ & $51.16^{+0.10}_{-0.05}$ & $-0.20^{+0.04}_{-0.05}$ & $0$ \\
        & 000418 & $-0.91^{+0.43}_{-0.56}$ & $53.71^{+0.70}_{-1.27}$ & $52.17^{+0.49}_{-0.61}$ & $53.64^{+0.77}_{-1.07}$ & $-1.60^{+1.05}_{-0.66}$ & $2$ \\
        & 000926 & $-0.64^{+0.39}_{-0.22}$ & $54.08^{+0.72}_{-0.72}$ & $52.66^{+0.59}_{-0.56}$ & $54.07^{+0.69}_{-0.72}$ & $-1.66^{+0.40}_{-0.66}$ & $2$ \\
        & 010222 & $-0.05^{+0.05}_{-0.07}$ & $52.68^{+0.26}_{-0.37}$ & $52.81^{+0.10}_{-0.43}$ & $52.93^{+0.31}_{-0.25}$ & $-0.31^{+0.08}_{-0.09}$ & $0$ \\
        & 030329 & $-0.34^{+0.07}_{-0.06}$ & $53.01^{+0.10}_{-0.18}$ & $52.22^{+0.01}_{-0.11}$ & $53.08^{+0.08}_{-0.18}$ & $-0.88^{+0.10}_{-0.07}$ & $0$ \\
        & 050820A & $-0.43^{+0.23}_{-0.08}$ & $51.93^{+0.54}_{-0.35}$ & $52.81^{+0.48}_{-0.39}$ & $52.88^{+0.47}_{-0.39}$ & $-0.07^{+0.02}_{-0.02}$ & $2$ \\
        & 050904 & $-0.67^{+0.19}_{-0.25}$ & $51.28^{+0.14}_{-0.39}$ & $51.76^{+0.24}_{-0.09}$ & $51.85^{+0.22}_{-0.10}$ & $-0.06^{+0.02}_{-0.05}$ & $0$ \\
        & 060418 & $-0.66^{+0.10}_{-0.11}$ & $50.92^{+0.22}_{-0.17}$ & $51.16^{+0.12}_{-0.14}$ & $51.32^{+0.21}_{-0.12}$ & $-0.20^{+0.05}_{-0.07}$ & $0$ \\
        & 090328 & $-0.85^{+0.58}_{-0.18}$ & $51.58^{+0.50}_{-0.63}$ & $51.45^{+0.42}_{-0.24}$ & $51.99^{+0.30}_{-0.47}$ & $-0.11^{+0.10}_{-0.36}$ & $0$ \\
        & 090423 & $-0.52^{+0.20}_{-0.32}$ & $53.09^{+0.70}_{-0.61}$ & $52.82^{+0.17}_{-0.47}$ & $53.19^{+0.55}_{-0.57}$ & $-0.50^{+0.27}_{-0.43}$ & $0$ \\
        & 090902B & $-0.83^{+0.34}_{-0.11}$ & $52.71^{+0.40}_{-0.38}$ & $54.36^{+0.28}_{-0.55}$ & $54.38^{+0.24}_{-0.57}$ & $-0.01^{+0.01}_{-0.02}$ & $2$ \\
        & 090926A & $-0.25^{+0.25}_{-0.73}$ & $51.09^{+1.25}_{-0.27}$ & $52.38^{+0.16}_{-0.20}$ & $52.41^{+0.37}_{-0.19}$ & $-0.05^{+0.04}_{-0.19}$ & $0$ \\
        & 120521C & $-0.47^{+0.12}_{-0.14}$ & $50.91^{+0.21}_{-0.23}$ & $50.79^{+0.21}_{-0.24}$ & $51.16^{+0.21}_{-0.20}$ & $-0.32^{+0.09}_{-0.12}$ & $0$ \\
        & 130427A & $-0.12^{+0.11}_{-0.32}$ & $51.91^{+0.65}_{-0.34}$ & $53.14^{+0.49}_{-0.29}$ & $53.17^{+0.49}_{-0.30}$ & $-0.03^{+0.02}_{-0.04}$ & $2$ \\
        & 130702A & - & $51.08^{+1.03}_{-0.84}$ & $49.08^{+0.23}_{-0.12}$ & $51.07^{+1.03}_{-0.79}$ & $-2.25^{+0.99}_{-0.86}$ & $0$ \\
        & 130907A & - & $50.23^{+0.04}_{-0.06}$ & $51.45^{+0.34}_{-0.11}$ & $51.47^{+0.33}_{-0.10}$ & $-0.01^{+0.00}_{-0.01}$ & $0$ \\
        & 140304A & $-0.73^{+0.12}_{-0.12}$ & $51.63^{+0.23}_{-0.19}$ & $50.19^{+0.46}_{-0.13}$ & $51.65^{+0.23}_{-0.17}$ & $-1.39^{+0.38}_{-0.28}$ & $2$ \\

        \hline
        \hline
    \end{tabular}
    \begin{flushleft}
    \begin{small}
    Note: The mode of the distribution of $E_\mathrm{true}$ is smaller than the mode of the $E_{\gamma, \mathrm{true}}$ for GRB 980703, and it is smaller than $E_\mathrm{K, true}$ for GRBs 000418, 000926 and 130702A. This is not due to an error in the analysis, but rather a combined effect due to addition in linear space whilst the distribution is in logarithmic space, and uncertainty in the Gaussian kernel estimator when determining the mode. In any case, the differences are well within the reported uncertainties and do not affect any results.
    \end{small}
    \end{flushleft}
\end{table*}

\begin{figure}
 \includegraphics[width=\columnwidth]{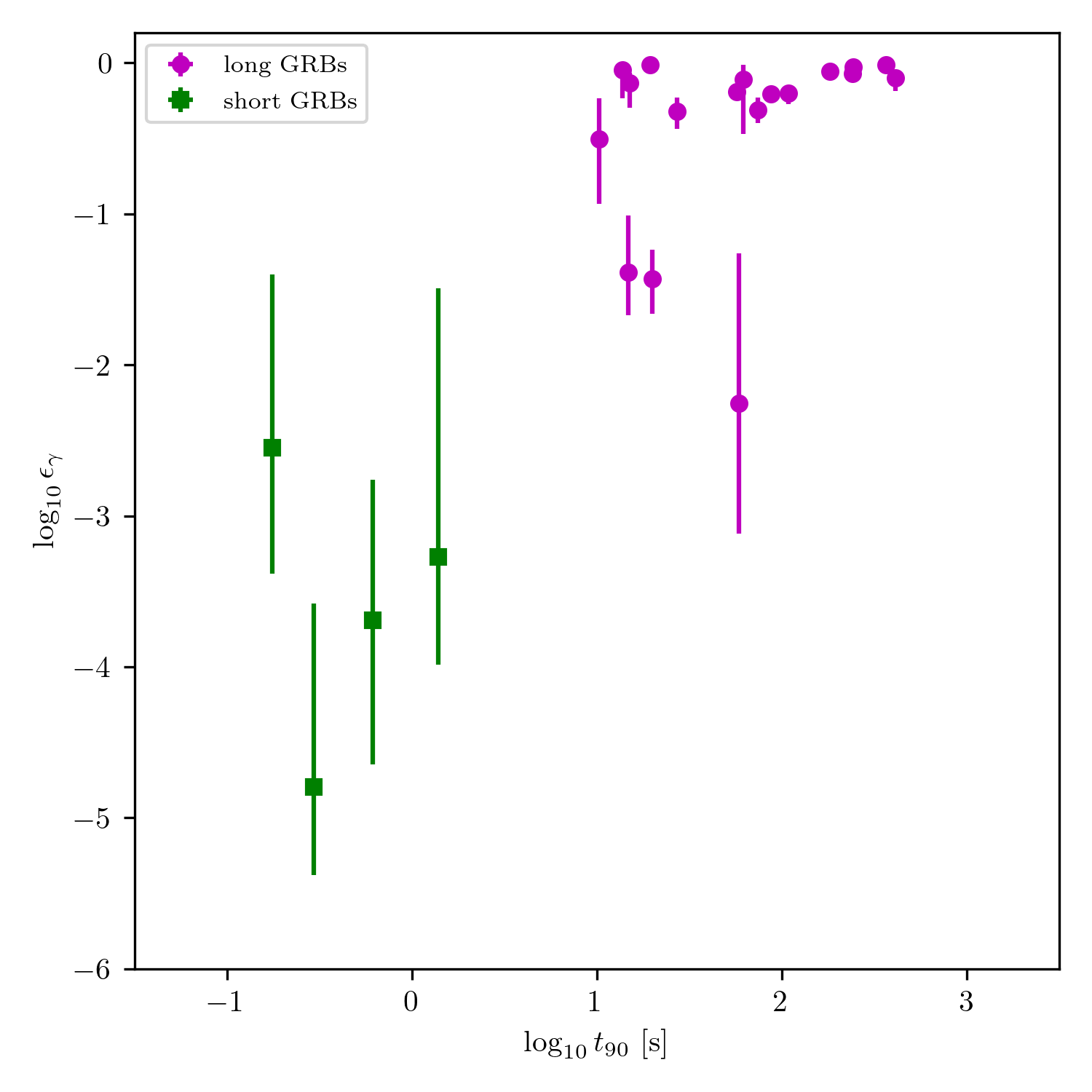}
 \caption{Scatter plot of the reported $t_{90}$ and derived $\epsilon_\gamma$ parameters for the short and long GRB sample. The magenta circles and green squares represent long and short GRBs, respectively. Error bars represent the 68\% credible interval.}
 \label{fig:plt-hist-epsilon_gamma-type}
\end{figure}

We present the relevant derived parameters of the GRB sample in Table~\ref{tab:derived_parameters}. 
In Figure \ref{fig:plt-hist-epsilon_gamma-type} it can be seen that short GRBs are systematically less efficient than long GRBs. The average value of the inferred $\epsilon_\gamma$ parameters for short and long GRBs are $2.7\times10^{-4}$ and $0.26$, respectively. \cite{2015MNRAS.454.1073B} have analyzed the observed flux from 10 long GRBs in X-rays and GeV energies to estimate the energetics of these GRBs. They find that there is a discrepancy between the estimated energies using both bands, where the energy estimated from the GeV flux is significantly larger. They state that this discrepancy can be explained within the forward shock framework by either assuming that the cooling break lies between these two bands or by taking into account Compton cooling effects. As a result they find that the average prompt efficiency value becomes $0.87$ for the X-ray estimated energies, and $0.14$ for GeV estimated energies. They note that the GeV estimated energetics should be more reliable, and these are consistent with our results. We perform a KS test to determine if the $\epsilon_\gamma$ distribution for long and short GRBs originate from a common distribution. We find that the null hypothesis can be significantly rejected, with $p=1.3\times10^{-4}$, so short GRBs
are indeed less efficient $\gamma$-ray emitters. However, this analysis does not account for the uncertainties in the $\epsilon_\gamma$ parameter, and relies only on the mode of the posterior distribution.

It is not so clear what might cause this difference in efficiency, and our study does not speak much to this
because we examine only the physics of the afterglow. We do note that in the afterglow, the short GRB blast waves have
a lower (synchrotron) emission efficiency, because this scales as $\epsilon_e\epsilon_B$, which is lower on average
for short GRBs (which have about the same $\epsilon_e$ and lower $\epsilon_B$ for a given true energy -- see below).
Whether we should expect that same difference to exist for the internal shocks that cause the prompt emission (or 
whether those are even dominated by synchrotron emission) is unclear, however.
In \cite{2019MNRAS.488.1416G}, the authors find that the main factor determining the radiative efficiency is the amount of baryon loading. Since our findings indicate that short GRBs are less efficient, this could also mean that baryon loading in short GRB jets is more prominent. \cite{2021arXiv210200005G} \citep[see also][]{2020MNRAS.495..570G, 2021MNRAS.500.3511G} have performed RMHD simulations to investigate how the prompt emission features (variability, spectrum and efficiency) vary for hydrodynamic and magnetized jets for both intermittent and continuous central engine activity scenarios. They have found that magnetized + intermittent jets are the most likely candidate for GRBs, as they yield high prompt efficiencies and are consistent with observed spectral and temporal features. They also note that, for hydrodynamic + intermittent jets, the efficiency drops below 1\%. As the degree of magnetization increases the mixing processes become less efficient and lead to higher $\epsilon_\gamma$ values. 

In Tables~\ref{tab:parameters} and \ref{tab:derived_parameters} it can be seen that, in some cases, the derived values for the microphysical parameters are difficult to reconcile with our theoretical understanding (i.e., $\epsilon_B + \epsilon_e \sim 1$). However, as mentioned previously, due to the $\xi_N$ degeneracy these values can be scaled down to more reasonable values by assuming a smaller value for $\xi_N$. This is supported by particle-in-cell simulations which indicate that $\xi_N$ can be as small as 0.01 \citep{Sironi2011}. We find that the derived values for the prompt efficiencies, $\epsilon_\gamma$, is approaching unity in some cases, which can be resolved by once again assuming a lower value for $\xi_N$ to scale down the values for the prompt efficiency. Moreover, even if we lower the $\xi_N$ value, assuming that the $\epsilon_e$ value is the same for both the prompt and afterglow phases, some of the modelling results suggest $\epsilon_\gamma > \epsilon_e$. Such high efficiencies are consistent with the proposed scenario by \cite{2001ApJ...551..934K}.

\cite{2018ApJ...859..160W} have analysed the optical and X-ray light curves of a large sample of GRBs to determine any achromatic jet-breaks present in the data sets. They infer the energetics and the opening angle of the GRB sample by assuming fiducial values for the other GRB parameters and by making use of analytical prescriptions. They find that the energetics are distributed as $\log_{10}E_{K, \mathrm{iso}}=54.82\pm0.56$ and $\log_{10}E_{K, \mathrm{true}}=51.33 \pm 0.58$ for the long GRB population. In our analysis we find $\log_{10}E_{K, \mathrm{iso}}=53.38\pm0.72$ and $\log_{10}E_{K, \mathrm{true}}=51.81 \pm 1.08$ for the long GRB population. We find that the energetics exhibit a wider spread across the long GRB population. We also find lower values for the mean of the $E_{K, \mathrm{iso}}$ distribution. Since \cite{2018ApJ...859..160W} only consider GRBs with an observed achromatic jet-break, their selection effects are different than our sample.

 In order to test whether or not the inferred energetics for the GRBs are feasible, we make use of the total beaming corrected energies, $E_{\mathrm{true}} \equiv E_{K, \mathrm{true}} + E_{\gamma, \mathrm{true}}$. Assuming that the jet is powered by the rotational energy of a Kerr black hole (e.g. \citealt{1977MNRAS.179..433B}), it is possible to estimate the total jet energy from the mass $M_\mathrm{BH}$ and rotation parameter $a$ of the black hole as
\begin{equation}
    E_\mathrm{true} = \epsilon_\mathrm{jet} E_\mathrm{rot} = 
                      \epsilon_\mathrm{jet} f(a) M_{\mathrm{BH}} c^2,
    \label{eq:M_BH}
\end{equation}
where $E_\mathrm{rot}$ is the rotational energy of the central black hole. Taking not too aggressive 
values $a=0.9$ and $\epsilon_\mathrm{jet}=0.1$ (e.g., \citealt{2000PhR...325...83L, 2005ApJ...630L...5M}), we get $E_\mathrm{true}\simeq 0.015M_\mathrm{BH}c^2$. 
The short and long GRBs with the highest inferred beaming corrected energies are GRBs 200522A and 090902B, respectively, with best-fit values of $2.2\times10^{52}$ and $2.4\times10^{54}$ erg. Using Equation~\ref{eq:M_BH}, the implied mass of the central black hole can be inferred as $>0.19$ and $>23.73$ $M_\odot$ for GRBs 200522A and 090902B, neither of which presents a significant difficulty for the favourite source models.

\subsection{$\epsilon_B$--$E_{K, \mathrm{true}}$, $\theta_0$--$\epsilon_B$ anti-correlations\label{sec:discsho}}

% \multirow{2}{*}{\multicolumn{1}{l}{Anti-/correlation}}
\begin{table}
    \setlength{\tabcolsep}{5pt}
    \def\arraystretch{1.25}
    \caption{Correlation significance results for the Jackknife re-sampling. The upper panel shows $p$-value estimates for the whole long GRB sample, the middle panel takes into account only the long GRBs in ISM-like environments, and the bottom panel is for the long GRBs in wind-like environments.}
    \begin{tabular}{|c|l|r@{.}l|r@{.}l|r@{.}l|}
    \hline
    \multicolumn{1}{l}{} & \multirow{2}{*}{Anti-/correlation} & \multicolumn{6}{c}{$p$-value} \\
    \multicolumn{1}{l}{} & \multicolumn{1}{l}{} & \multicolumn{2}{r}{Minimum} & \multicolumn{2}{r}{Maximum} & \multicolumn{2}{r}{Average}\\
    \hline
    \hline
    \parbox[t]{2mm}{\multirow{3}{*}{\rotatebox[origin=c]{90}{long GRBs}}} & $\epsilon_B\mathrm{-}E_{K, \mathrm{true}}$ & $1$&$22\times10^{-7}$ & $2$&$28\times10^{-4}$ & $4$&$65\times10^{-5}$ \\ 
    & $\theta_0\mathrm{-}\epsilon_B$ & $6$&$85\times10^{-4}$ & $0$&$010$ & $4$&$18\times10^{-3}$ \\
    & $\epsilon_B\mathrm{-}E_{K, \mathrm{iso}}$ & $0$&$011$ & $0$&$48$ & $0$&$12$  \\
    % & $\theta_0\mathrm{-}E_{K, \mathrm{true}}$ & $6$&$08\times10^{-6}$ & $4$&$92\times10^{-5}$ & $2$&$10\times10^{-5}$  \\
    \hline
    \parbox[t]{2mm}{\multirow{3}{*}{\rotatebox[origin=c]{90}{ISM-like}}} & $\epsilon_B\mathrm{-}E_{K, \mathrm{true}}$ & $4$&$74\times10^{-6}$ & $1$&$28\times10^{-4}$ & $3$&$66\times10^{-5}$ \\ 
    & $\theta_0\mathrm{-}\epsilon_B$ & $1$&$71\times10^{-4}$ & $6$&$12\times10^{-3}$ & $1$&$60\times10^{-3}$ \\
    & $\epsilon_B\mathrm{-}E_{K, \mathrm{iso}}$ & $0$&$10$ & $0$&$99$ & $0$&$82$  \\
    % & $\theta_0\mathrm{-}E_{K, \mathrm{true}}$ & $2$&$34\times10^{-5}$ & $1$&$58\times10^{-3}$ & $4$&$36\times10^{-4}$  \\
    \hline
    \parbox[t]{2mm}{\multirow{3}{*}{\rotatebox[origin=c]{90}{Wind-like}}} & $\epsilon_B\mathrm{-}E_{K, \mathrm{true}}$ & $0$&$011$ & $0$&$68$ & $0$&$21$ \\ 
    & $\theta_0\mathrm{-}\epsilon_B$ & $0$&$55$ & $0$&$99$ & $0$&$87$ \\
    & $\epsilon_B\mathrm{-}E_{K, \mathrm{iso}}$ & $9$&$03\times10^{-3}$ & $0$&$37$ & $0$&$12$  \\
    % & $\theta_0\mathrm{-}E_{K, \mathrm{true}}$ & $0$&$15$ & $0$&$31$ & $0$&$20$  \\
    \hline
    \end{tabular}
    \label{tab:jackknife}
\end{table}

\begin{figure}
 \includegraphics[width=\columnwidth]{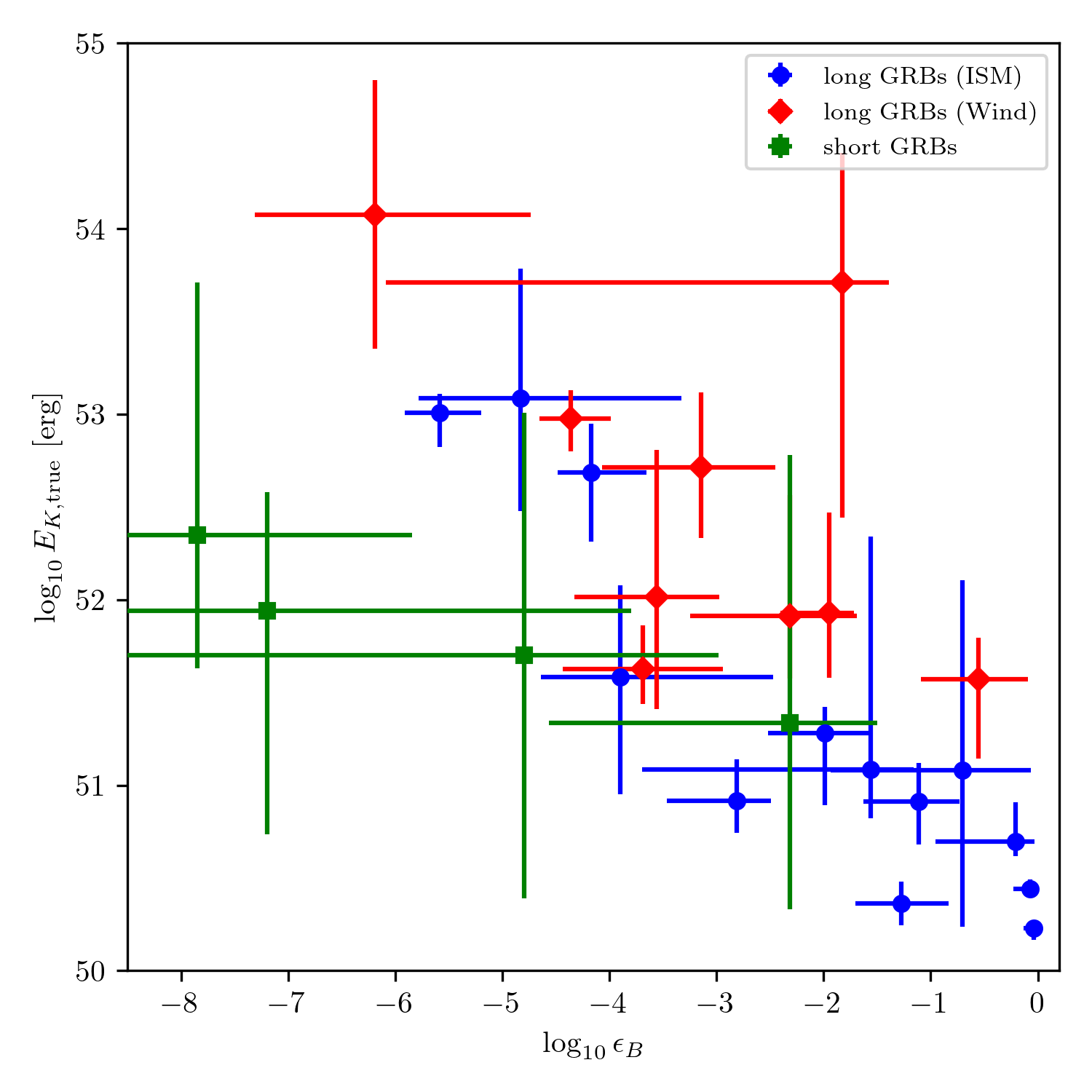}
 \caption{Scatter plot of $\epsilon_B$ and $E_{K, \mathrm{true}}$ parameters for the GRB sample. The blue circles, red diamonds, and green squares represent ISM-like long GRBs, wind-like long GRBs and short GRBs, respectively. Error bars represent the 68\% credible interval.}
 \label{fig:plt-corner-epsilon_B-E_K_true}
\end{figure}

We find that $\theta_0$ and $E_{K, \mathrm{true}}$ are strongly correlated with each other for both wind-like and
ISM-like long GRBs; this does not, however,
have any new meaning. It is simply the result of the fact that $E_\mathrm{true}$ is derived from $E_\mathrm{iso}$ via
the opening angle, and that the distribution of $E_\mathrm{iso}$ is fairly narrow, whereas that of $\theta_0$ is wider.
Hence, this correlation is largely due to the fact that we are correlating $\theta_0$ with itself.

We do however find a strong and significant anti-correlation between $\epsilon_B$ and $E_{K, \mathrm{true}}$ for the sample of long GRBs. In Figure~\ref{fig:plt-corner-epsilon_B-E_K_true} we demonstrate this anti-correlation. We find that the fraction of energy lost to amplifying magnetic fields systematically decreases as the measured beaming corrected kinetic energy gets larger.  When we perform a Pearson $r$ correlation test we find a $p$-value of $10^{-5}$,
strongly rejecting the null hypothesis that they are uncorrelated.
Alternatively, we can of course regard this correlation as due to an $\epsilon_B-\theta_0$ relation. We also checked for a possible correlation between $\epsilon_B$ and $E_{K, \mathrm{iso}}$; it is not significant. This may be somewhat surprising, since we think of the energy
fractions in electrons and magnetic field to be set very locally at the shock,
and thus correlate better with $E_{K, \mathrm{iso}}$, which scales with the
local energy per unit area at the shock.

Since the number of GRBs in either sample is small, we use a Jackknife resampling test to check the robustness of
these correlations due to outliers. The results can be seen in Table~\ref{tab:jackknife}, and show that the significant correlations are robust. However, they also reveal are rather strong difference between the two subclasses: the correlations are not significant in the wind-like GRBs and very strongly significant in the ISM-likes. The significance
of the result for the total population is therefore entirely due to that of the ISM-like GRBs. This is puzzling, since it unclear how the correlation between these two blast wave parameters would come to depend on the shape of the ambient density distribution.

\subsection{Caveats}

First of all, it is not straightforward to infer population distributions from uncertain measurements. When creating histograms we only considered the point estimate (i.e., mode of the posterior distribution) of the inferred parameter values. This approach does not take into account the full information contained in the obtained parameter distributions. It would be valuable to combine the posterior distributions from individual modelling efforts to estimate how the parameters are distributed across the GRB population in a statistically correct manner (e.g., \citealt{2010PhRvD..81h4029M, 2010ApJ...725.2166H}), however this is out of the scope of this work.

Second, it is not possible to compile an unbiased sample of well-sampled afterglows, because a variety of instrumental biases and observer choices enter into the determination of which GRBs to follow up extensively and for which such followup is successful. Therefore, the inferred GRB population will likely not cover all of the physical parameter space. We have commented above on whether we estimate this has a significant influence on our conclusions. While we accounted for the small sample size of especially short GRBs when stating significances, it is still good to bear in mind that our short GRBs are especially unrepresentative of the total population of short GRBs (though, as we noted, there is no clear expectation that this this would bias the intrinsic properties of this subset of short GRBs, for which we draw
the most marked conclusions.)

\section{Conclusion}
\label{sec:conclusion}

We have studied a sample of 26 GRB afterglows (as well as the total prompt $\gamma$-ray energy emitted),
which was biased to enabling detailed afterglow physics studies, i.e., towards having well-sampled radio,
optical, and X-ray light curves. While this largely excludes the most obscured GRBs (due to optical extinction) and GRBs in low-density regions (i.e., most short GRBs), we argue that there are quite a few
conclusions about GRB physics that are not strongly affected by those biases:
\begin{enumerate}
    \item All physical parameters have intrinsic distributions of significant width, i.e., none have a
          `standard' value that is almost the same for all GRBs, or even within a subsample (short, long-ISM, or long-wind; Sect.~\ref{sec:results}).
    \item Short GRBs prefer uniform ambient densities, in agreement with theoretical expectations and  
          previous studies (Sect.~\ref{sec:resenv}).
    \item Long GRBs have about equal likelihood of wind-like and uniform ambient media. 
    %The uniform media are not easily understood in the context of their origin in massive stars with strong winds (Sect.~\ref{sec:discenv}).
          A massive star progenitor is expected to impact the environment of the burst, suggesting a wind-like medium to be more likely. We note that even a massive star wind environment can be close to homogeneous at scales probed by the afterglow observations, but that this is not the most natural outcome for typical parameters (Sect.~\ref{sec:discenv}).
    \item The wind strengths for the wind-like long GRBs favour canonical mass loss parameters of
          massive Wolf-Rayet stars, the most likely progenitors, and specifically do not indicate a
          bias towards low mass loss rates, as required by some GRB models (Sect.~\ref{sec:discenv}).
    \item We do not find evidence for different jet opening angles between long and short GRBs (Sect.~\ref{sec:resene}).
    \item The observer viewing angles are consistent with top-hat jets, and with the opening angles of the 
          prompt $\gamma$-ray emission and early afterglow emission being the same.
    \item We find a distribution of slopes of the energy distribution of accelerated electrons, $p$, that
          is consistent with previous studies; it contains only a few examples where $p<2$ significantly,
          but many where it is close enough to 2 to warrant caution in fitting $p$ and $\epsilon_e$ (Sect.~\ref{sec:ressho}).
    \item The values of $\epsilon_e$ are all in the range 0.1\,--\,1, with no significant differences     
          between short/long or wind/uniform samples (Sect.~\ref{sec:ressho}).
    \item The true total energies of long and short GRBs are similar, implying that the relative faintness
          of short GRBs in $\gamma$ rays is due to their lower $\gamma$-ray emission efficiency (Sect.~\ref{sec:discene}).
    \item Some required gamma-ray efficiencies of GRBs are close to 1, 
          which is a challenging value for current prompt emission
          theories
          (Sect.~\ref{sec:discene}).
    \item There is a strong and significant correlation for ISM-like long       GRBs between the magnetic field energy at the shock and the 
          true total kinetic energy of the blast
          wave. It is surprising that this same correlation does not exist for the wind-like GRBs (Sect.~\ref{sec:discsho}).
\end{enumerate}

\section*{Acknowledgements}
We thank the anonymous referee for their insightful comments. This work was carried out on the Dutch national e-infrastructure with the support of SURF Cooperative. H. J. van Eerten acknowledges partial
support by the European Union Horizon 2020 Programme under the
AHEAD2020 project (grant agreement number 871158).

\section*{Data Availability}
A reproduction package including the data underlying this article will be made available at Zenodo with DOI: 10.5281/zenodo.5035173.

%%%%%%%%%%%%%%%%%%%%%%%%%%%%%%%%%%%%%%%%%%%%%%%%%%

%%%%%%%%%%%%%%%%%%%% REFERENCES %%%%%%%%%%%%%%%%%%

% The best way to enter references is to use BibTeX:

\bibliographystyle{mnras}
\bibliography{references} % if your bibtex file is called example.bib

%%%%%%%%%%%%%%%%%%%%%%%%%%%%%%%%%%%%%%%%%%%%%%%%%%

%%%%%%%%%%%%%%%%% APPENDICES %%%%%%%%%%%%%%%%%%%%%

\appendix

\section{Posterior distributions for the free physical parameters}
\begin{figure*}
 \includegraphics[width=\textwidth]{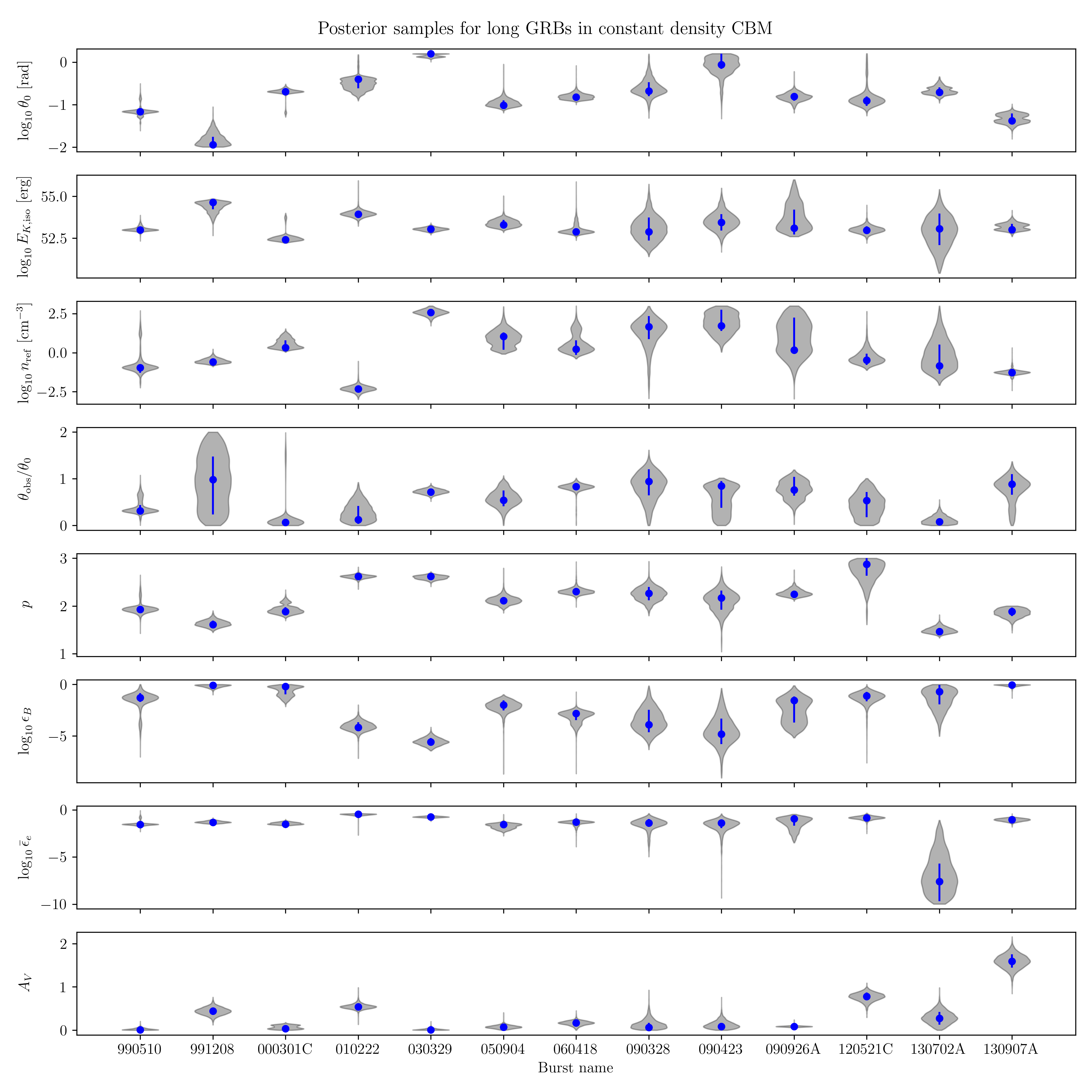}
 \caption{Violin plots representing the obtained posterior distributions of the free physical parameters for the long GRB sample associated with a constant density environment. The shaded areas represent the density of the posterior samples and circles represent the mean value. Error bars represent the 68\% credible interval.}
 \label{fig:plt-parameters-ism}
\end{figure*}
\begin{figure*}
 \includegraphics[width=\textwidth]{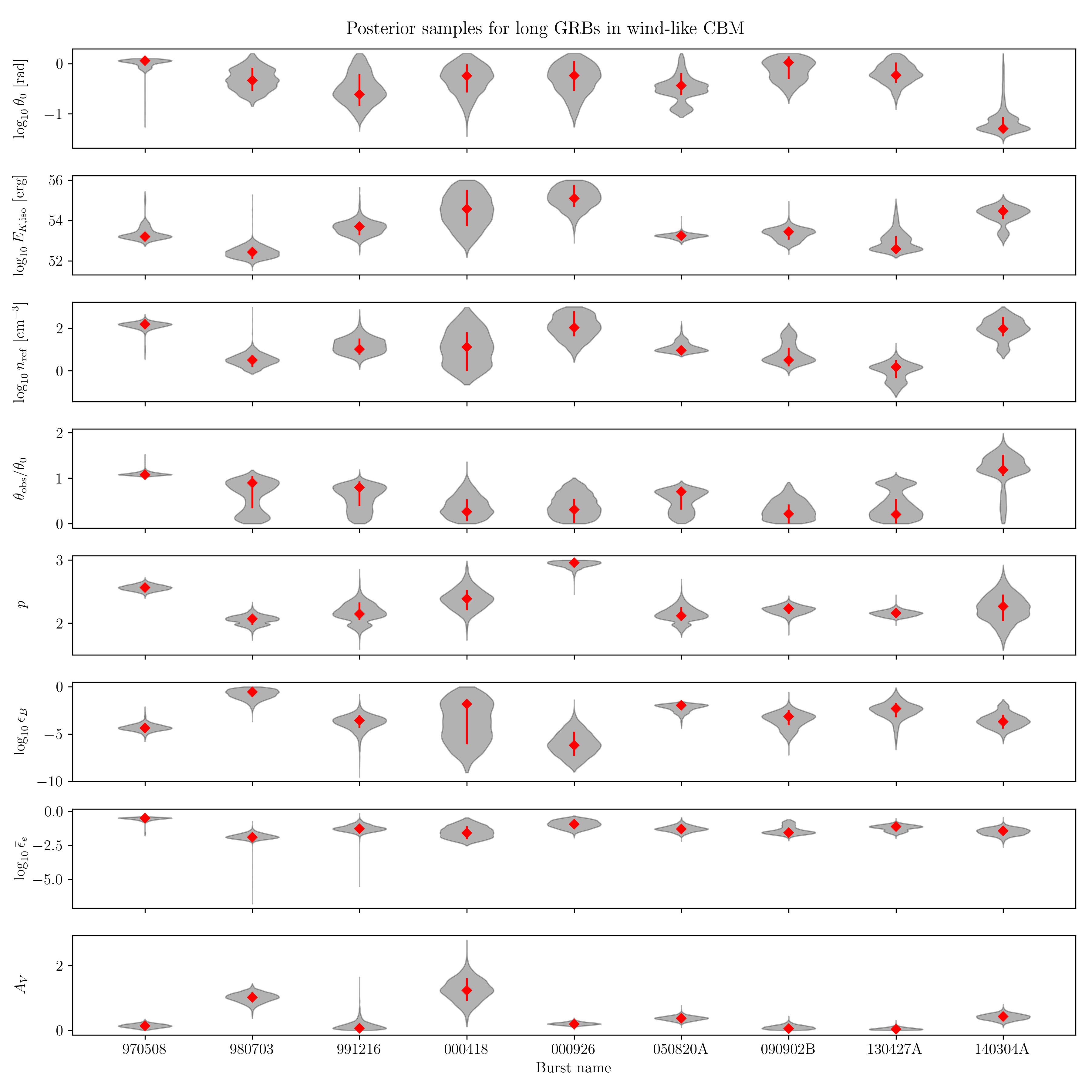}
 \caption{Violin plots representing the obtained posterior distributions of the free physical parameters for the long GRB sample associated with a wind-like environment. The shaded areas represent the density of the posterior samples and diamonds represent the mean value. Error bars represent the 68\% credible interval.}
 \label{fig:plt-parameters-wind}
\end{figure*}
\begin{figure}
 \includegraphics[width=.8\columnwidth]{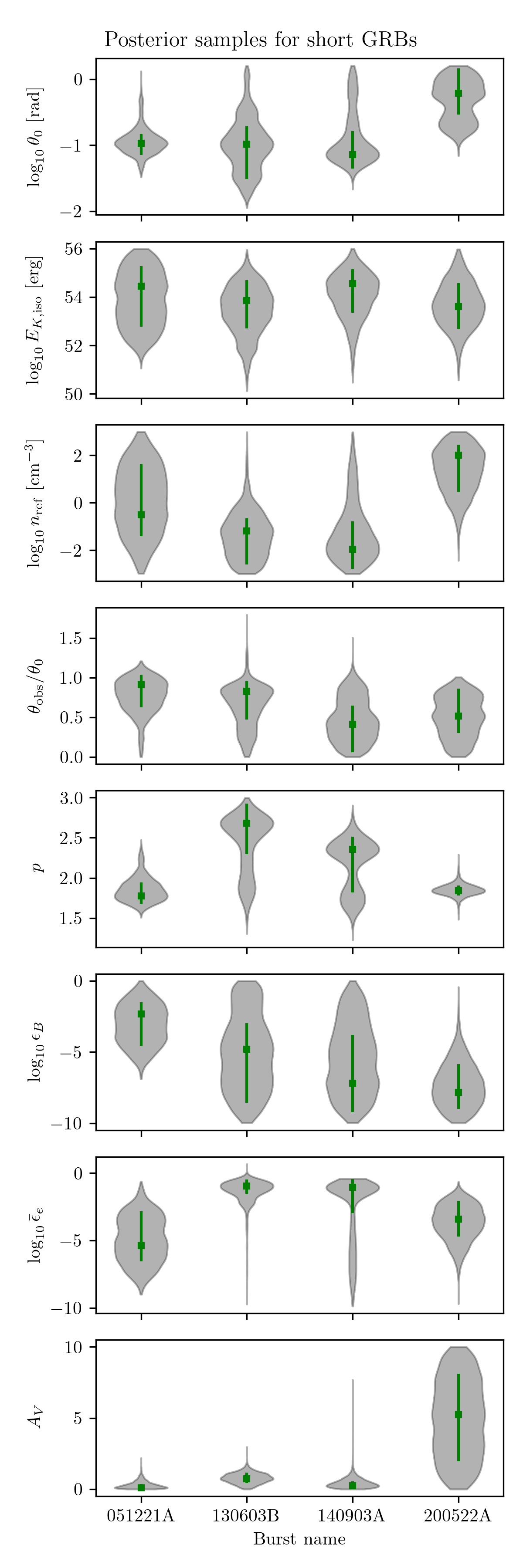}
 \caption{Violin plots representing the obtained posterior distributions of the free physical parameters for the short GRB sample. The shaded areas represent the density of the posterior samples and squares represent the mean value. Error bars represent the 68\% credible interval.}
 \label{fig:plt-parameters-sgrb}
\end{figure}

%%%%%%%%%%%%%%%%%%%%%%%%%%%%%%%%%%%%%%%%%%%%%%%%%%

% Don't change these lines
\bsp	% typesetting comment
\label{lastpage}
\end{document}